\documentclass[]{resources/aa}
\usepackage{graphicx}
\usepackage[varg]{txfonts}
\usepackage{physics}
\usepackage{amsmath, amssymb}
\usepackage{dsfont}
\usepackage{bbm}
\usepackage[table]{xcolor}
\usepackage{ulem}
\usepackage{xpatch}
\usepackage{hyperref}
\usepackage{import}
\usepackage{makeidx}
\usepackage{changes}
\usepackage{placeins}

\newcommand{\sref}[1]{Sec.~\ref{#1}}
\newcommand{\tab}[1]{Table~\ref{#1}}
\newcommand{\fig}[1]{Fig.~\ref{#1}}
\newcommand{\app}[1]{Appendix~\ref{#1}}

\newcommand{\equ}[1]{Eq.~(\ref{#1})}
\newcommand{\equo}[1]{Eq.~\ref{#1}}

\newcommand{\equos}[2]{Eqs.~\ref{#1}~-~\ref{#2}}
\newcommand{\Msolpyr}{\mathrm{M_\odot~yr^{-1}}}
\newcommand{\Msol}{\mathrm{M_\odot}}
\newcommand{\Mdot}{\Dot{M}_\star}

\newcommand{\colout}[1]{\bgroup\markoverwith{\textcolor{#1}{\rule[.5ex]{2pt}{0.4pt}}}\ULon}

\makeatletter
\renewcommand*\aa@pageof{, page \thepage{} of \pageref*{LastPage}}

\makeatother

\makeindex

\begin{document}


%
\titlerunning{What governs the spin distribution of very young < 1 Myr low mass stars}
\authorrunning{L.~Gehrig \& E.~I.~Vorobyov}
\title{What governs the spin distribution of very young < 1 Myr low mass stars} 
%
%
\author{
 L.~Gehrig\inst{1} and
 E.~I.~Vorobyov\inst{1,2}
}
\institute{
 Department of Astrophysics, University of Vienna,
 Türkenschanzstrasse 17, A-1180 Vienna, Austria
 \and 
 Research Institute of Physics, Southern Federal University, Rostov-on-Don 344090, Russia 
}
\date{Received ....; accepted ....}

\abstract
{The origin of the stellar spin distribution at young ages is still unclear. Even in very young clusters ($\sim 1$~Myr), a significant spread is observed in rotational periods ranging from $\lesssim 1$ to $\sim 10$~days.}
{We aim to study the parameters that might govern the spin distribution of low mass ($\lesssim1.0~\mathrm{M_\odot}$) stars during the first million years of their evolution.}
{
We compute the evolution and rotational periods of young stars, using the MESA code, starting from a stellar seed, and take protostellar accretion, stellar winds, and the magnetic star-disk interaction into account.
Furthermore, we add a certain fraction of the energy of accreted material into the stellar interior as additional heat and combine the resulting effects on stellar evolution with the stellar spin model.
}
{
For different combinations of parameters, stellar periods at an age of 1~Myr range between 0.6~days and 12.9~days.
Thus, during the relatively short time period of 1~Myr, a significant amount of stellar angular momentum can already be removed by the interaction between the star and its accretion disk.
The amount of additional heat added into the stellar interior, the accretion history, and the presence of disk and stellar winds have the strongest impact on the stellar spin evolution during the first million years.
The slowest stellar rotations result from a combination of strong magnetic fields, a large amount of additional heat, and effective winds.
The fastest rotators combine weak magnetic fields and ineffective winds or result from a small amount of additional heat added to the star.
Scenarios that could lead to such configurations are discussed.
Different initial rotation periods of the stellar seed, on the other hand, quickly converges and do not affect the stellar period at all.
}
{
Our model matches up to 90\% of the observed rotation periods in six young ($\lesssim 3$~Myr) clusters.
Based on these intriguing results, we motivate to combine our model with a hydrodynamic disk evolution code to self-consistently include several important aspects such as episodic accretion events, magnetic disk winds, internal, and external photo-evaporation.
Such a combined model could replace the widely-used disk-locking model during the lifetime of the accretion disk and provide valuable insights into the origin of the rotational period distribution of young clusters.
}


\maketitle


\section{Introduction}
\label{sec:intro}

The origin of the stellar spin  and angular momentum (AM) distribution of young stars still poses problems for stellar evolution models. 
Observations of stellar rotation in young clusters over the last decade have provided valuable insights for the evolution of the stellar spin and AM \citep[e.g.,][]{herbst02, Rebull2006, Irwin08, Irwin09, Hartmann09, Affer13, Gallet13, Rebull2020, Serna2021}.
Especially the early distribution ($\lesssim10$~Myr) of stellar spin and AM is an important benchmark for gyrochronology studies \citep[e.g.,][]{Barnes07} and can affect stellar high energy radiation of low mass stars \citep[e.g.,][]{Pallavicini81, Micela85, Wright11, France18}.

We want to focus on the first Myr of stellar evolution, in which the star is embedded in an envelope and surrounded by an accretion disk.
During the accretion disk phase, it is assumed that the accretion disk plays an important role in regulating the stellar spin distribution.
Mass and AM can be exchanged between the star and the disk via magnetic field lines, influencing the stellar rotation period \citep[e.g.,][]{Armitage96, matt05, matt2010}.
Early models propose that the star rotates at the orbital period of the inner disk boundary \citep[disk-locking, e.g.,][]{Ghosh79, Koenigl91}.
However, because of a limited region, in which the magnetic field lines can connect to the accretion disk, the disk-locking mechanisms cannot remove enough AM from the star to counteract accretion and contraction and does not explain the existence of slowly rotating stars \citep[e.g.,][]{Matt05APSW, matt05, Zanni13}.
Since then additional mechanisms have been proposed that can remove AM from the star during the accretion disk phase: the process of accretion itself can drive accretion-powered stellar winds \citep[APSW,][]{Matt05APSW} and disk material can be ejection along magnetic field lines \citep[magnetospheric ejections, MEs, e.g.,][]{Zanni13} removing AM from the star.

Even at a very young age of $\sim 1$~Myr, young clusters already show a significant spread in stellar rotation periods.
Observations in the Orion Nebular Cluster (ONC) \citep[e.g.,][]{herbst02,Serna2021}, NGC~2264 \citep[e.g.,][]{Venuti2017}, Taurus \citep[][]{Rebull2020}, NGC~6530 \citep[][]{Henderson12}, and other young clusters \citep[YSOVAR,][]{Rebull2014} show rotation periods ranging from $<1$ to $\sim 10$~days. 
We note that there is a small number of stars rotating even faster or slower than this range at a young age \citep[e.g.,][]{Rebull2014}.
To date, the parameters responsible for this large spread are not constrained.
\cite{Bouvier2014} argues that processes during the cloud collapse or the embedded phase might govern the rotational distribution at $\sim 1$~Myr.
Later studies have not solved for the very early stellar spin evolution and adopted the observed spread from $\sim 1$~day to $\sim 10$~days as initial conditions \citep[e.g.,][]{Gallet19, Roquette21}.

To explore the parameter and processes responsible for the rotational distribution of low-mass stars ($\lesssim 1~\Msol$), we combine \textit{Modules for Experiments in Stellar Astrophysics} \citep[MESA][]{Paxton2011, Paxton2013, Paxton2015, Paxton2018, Paxton2019} with a recent stellar spin evolution model presented by \cite{Ireland21}.
Similar to current stellar evolution models, we start with a stellar seed ($M_\mathrm{\star} = 0.05~\mathrm{M_\odot}$) that accretes mass from the surrounding disk \citep[e.g.,][]{Baraffe2012, Vorobyov17c, Kunitomo17, Steindl21}.
A fraction of the gravitational energy of the accreted material is deposited into the stellar interior as additional heat, which can lead to an inflation of the stellar radius and affect the stellar evolution significantly \citep[e.g.,][]{Vorobyov17c, Kunitomo17}.
The amount of additional heat depends on the fraction of energy deposited in the stellar interior, the region, in which the energy is deposited, and the accretion rate onto the star.
The effect of those parameters on the stellar spin evolution, however, is not clear and will be tested in this study.
Our model considers the influence of stellar inflation on the spin evolution of stars. It also takes into account different rotation rates of the stellar seed for the spin evolution of these stars.
Furthermore, we can test a wide parameter space including the stellar magnetic field strength, the initial angular velocity distribution of the stellar seed, winds originating from the star and the disk, and the initial radii of the stellar seed.

Our results suggest that certain mechanisms and parameters have a major impact on the early stellar spin evolution, while others are virtually unimportant.
Distinct slow and fast rotating models differ in the rotation period at an age of 1~Myr by up to two orders of magnitude.
A comparison with observed rotational periods of six young clusters allows us to test these model results.
This study is structured as follows: 
\sref{sec:model_description} summarizes the model highlights and defines the parameter space used in our simulations.
The results are presented in \sref{sec:results}, discussed, and compared to observations in \sref{sec:discussion}.
Finally, we summarize our conclusions in \sref{sec:conclusion}.


\section{Model and parameter space}
\label{sec:model_description}

In this study, we combine version v12778 of MESA with the stellar spin evolution model of \cite{Ireland21}. In the following, we highlight important aspects of the respective models and provide the respective references for additional information. 
The adopted micro-physics and the treatment of rotation in MESA are summarized in \app{app:mesaphysics} and \app{app:mesarot}, respectively.
The MESA input files for re-creating our results are publicly available\footnote{will be uploaded to ZENODO after acceptance}.

\subsection{Stellar evolution with MESA}

Similar to recent stellar evolution studies \citep[e.g.,][]{Kunitomo17, Steindl21}, we start our simulations with a stellar seed with an initial stellar mass of $M_\mathrm{init} = 0.05~\Msol$\footnote{We note that \cite{Kunitomo17} and \cite{Steindl21} start with $M_\mathrm{init} = 0.01~\Msol$. The combination of high accretion rates and fast stellar rotation leads to convergence issues with MESA for stellar masses $<0.05~\Msol$}, solar metallicity, $Z_\star = 0.014$, and a Deuterium abundance of 20~ppm.
During the initial stages of stellar evolution, disk material is accreted onto the star. 
The associated energy of the accreted material is partly radiated away in the form of an accretion luminosity
\begin{equation}
    L_\mathrm{acc} = (1-\beta) \epsilon \frac{G M_\star \Dot{M}_\star}{R_\star}
\end{equation}
and partly added to the stellar interior as additional heat
\begin{equation}\label{eq:eadd}
    \Dot{E}_\mathrm{add} = \beta \epsilon \frac{G M_\star \Dot{M}_\star}{R_\star} \, ,
\end{equation}
with the stellar mass $M_\star$, the stellar radius $R_\star$, the gravitational constant $G$ and the accretion rate onto the star $\Dot{M}_\star$. The geometry of the accretion process is described by $\epsilon$. Accretion from a thin accretion disk is described with $\epsilon \leq 0.5$ \citep[][]{Hartmann97} and we choose $\epsilon = 0.5$ in this study \citep[e.g.,][]{Baraffe09,Steindl21}.
The amount of additional energy added to the star is controlled by $\beta$ and is assumed to depend on the accretion rate \citep[e.g.,][]{Baraffe2012, Vorobyov17c, Kunitomo17, Elbakyan2019, Steindl21}.
During phases of low accretion rates, the disk is truncated at several stellar radii by magnetic field lines anchored in the star \citep[e.g.,][]{bessolaz08,hartmann16}.
Disk material is channeled through so-called funnels and is accreted onto the star at nearly free-fall speeds.
Close to the stellar surface, a shock front develops, and most of the material's energy is radiated away \citep[e.g.,][]{Koenigl91}.
As a result, only a small fraction of the accreted energy is added to the stellar interior and $\beta$ is small.
During phases of high accretion rates, the disk material is pushed toward the star and the stellar magnetic field is no longer able to channel the disk material into the funnel flows. 
Without the radiative loss of energy at the shocks close to the stellar surface a larger amount of energy can be added to the star.
Based on previous studies, accretion rates of $\Mdot \gtrsim 10^{-5}~\Msolpyr$ result in high values of $\beta = \beta_\mathrm{up}$ and accretion rates of $\Mdot \lesssim 10^{-7}~\Msolpyr$ result in low values of $\beta = \beta_\mathrm{low}$ \citep[e.g.,][]{Baraffe2012,Vorobyov17c, Jensen2018}.
In between the value of $\beta$ transitions smoothly between $\beta_\mathrm{up}$ and $\beta_\mathrm{low}$ according to
\begin{equation}\label{eq:beta}
    \beta(\Mdot) = \frac{\beta_\mathrm{up}+\beta_\mathrm{low}}{2} + \frac{\beta_\mathrm{up}-\beta_\mathrm{low}}{2} \tanh{ \frac{\log{\Mdot}+6}{0.5} } \, ,
\end{equation}
with the accretion rates in units of $\Msolpyr$.
In this study, we choose $\beta_\mathrm{low}=0.005$ and vary $\beta_\mathrm{up}$ between 0.005 (corresponding to the case of cold accretion, $\beta=\mathrm{const.}=0.005$) and 0.5 (corresponding to the case of warm or hybrid accretion).
$\Dot{E}_\mathrm{add}$ is distributed linearly starting from the stellar surface, down to a relative mass coordinate $M_\mathrm{add}$ \citep[e.g.,][]{Kunitomo17, Steindl21}. We choose $M_\mathrm{add}$ to be in the range of $0.01-0.1$, meaning that $\Dot{E}_\mathrm{add}$ is distributed within the outer $1-10$\% of the stellar mass.
$\Dot{E}_\mathrm{add}$ can cause the star to inflate, depending on the value of $\beta$ and the position in the stellar interior, up to which this energy is added $M_\mathrm{add}$ \citep[e.g.,][]{Baraffe09, Baraffe2012, Vorobyov17c, Kunitomo17, Steindl21}.

The outer, atmospheric boundary condition is chosen according to \cite{Kunitomo17}. 
For stellar radii larger than $0.7~\mathrm{R_\odot}$, we choose the  “photospheric” table with an optical depth of $\tau_\mathrm{s}=2/3$, a surface temperature $T_\mathrm{s}=T_\mathrm{eff}$ and the surface pressure $P_\mathrm{s}$ obtained from PHOENIX \citep[][]{Hauschildt1999a, Hauschildt1999b} and ATLAS9 \citep[][]{Castelli2003} model atmospheres.
For smaller stellar radii, we adopt the “$\tau_\mathrm{s}=100$” model, in which $P_\mathrm{s}$ and $T_\mathrm{s}$ are calculated at $\tau_\mathrm{s}=100$ from ATLAS9 and COND \citep[][]{Allard2001} model atmospheres.

\subsection{Stellar spin evolution}\label{sec:spin_model}

In the scope of our model, the star rotates as a solid body with the stellar AM, $J_\star = I_\star \Omega_\star$; $I_\star$ and $\Omega_\star$ are the stellar moment of inertia and angular velocity $\Omega_\star$, respectively.
External torques $\Gamma_\mathrm{ext}$ change $J_\star$ according to $\Dot{J}_\star = I_\star \Dot{\Omega}_\star + \Dot{I}_\star \Omega_\star = \Gamma_\mathrm{ext}$.
This results in a stellar spin evolution according to 
\begin{equation}\label{eq:stellen_spin_evolution}
    \Dot{\Omega}_\star = \frac{\Gamma_\mathrm{ext}}{I_\star} - \frac{\Dot{I}_\star}{I_\star} \Omega_\star \, .
\end{equation}
The moment of inertia $I_\star$ is calculated by the MESA code and the external torque contributions $\Gamma_\mathrm{ext}$ are calculated according to the description provided by \cite{Ireland21}.
The contributions to $\Gamma_\mathrm{ext}$ is composed of the influence of accretion $\Gamma_\mathrm{acc}$, magnetospheric ejections $\Gamma_\mathrm{me}$ and stellar winds $\Gamma_\mathrm{w}$.
The internal contributions due to accretion, mass loss, and changes in the stellar radius are combined in $\Gamma_\mathrm{int} = -\Dot{I}_\star \Omega_\star$. 
An increase in stellar mass or the stellar radius results in a positive $\Dot{I}_\star$ and the star spins down, for example, during phases of stellar radius inflation.
On the other hand, the star spins up if the star contracts.
According to \cite{Ireland21}, the external torque contributions depend on different stellar parameters and disk parameters.
We want to highlight the general dependencies and specific trends of the individual torque values.
The applicability of the spin model presented in \cite{Ireland21} for this study is discussed in \sref{sec:applicability}.
For additional information, we refer to the original article.

\textit{Accretion torque $\Gamma_\mathrm{acc}$:}
Material that accretes onto the star usually adds the AM of the disk material at the inner disk boundary to the star \citep[e.g.,][]{Matt10, Gallet19, Ireland21}, increasing the stellar AM, and causing a stellar spin-up according to 
\begin{equation}\label{eq:gamma_acc}
    \Gamma_\mathrm{acc} = |\Mdot| (\Omega(R_\mathrm{t})- \Omega_\star) R_\mathrm{t}^2 \, ,
\end{equation}
with the disk rotation rate at the truncation radius $R_\mathrm{r}$, $\Omega(R_\mathrm{t})$\footnote{We note that the formulation of $\Gamma_\mathrm{acc}$ in \cite{Ireland21} do not include the $-\Omega_\star$ term. They take into account stellar rotation periods much smaller compared to the rotation rates at the inner disk boundaries. At young ages, however, fast rotation rates close to break-up should be considered and the additional term should be included.}. 
The truncation radius in units of the stellar radius can be written as
\begin{equation}
    \frac{R_\mathrm{t}}{R_\star} = K_\mathrm{t} \left( \frac{B_\star^2 R_\star^2}{4 \pi |\Mdot| v_\mathrm{esc}} \right)^{m_\mathrm{t}} \, ,
\end{equation}
with $K_\mathrm{t}=0.756$, $m_\mathrm{t}= 0.34$, the stellar dipole field strength $B_\star$, and the escape velocity from the stellar surface $v_\mathrm{esc} = (2 G M_\star / R_\star)^{1/2}$.
The disk at the truncation radius is assumed to rotate slower compared to the Keplerian velocity \citep[e.g.,][]{bessolaz08, Zanni13} due to the removal of AM from the disk due to the magnetic star-disk interaction \citep[e.g.,][]{Zanni13} or disk winds \citep[e.g.,][]{Zhang2017}.
Thus, $\Omega(R_\mathrm{t})$ can be written as
\begin{equation}\label{eq:non_kep}
    \Omega(R_\mathrm{t}) = K_\mathrm{acc} \left( \frac{R_\mathrm{t}}{R_\star} \right)^{m_\mathrm{acc}} \Omega_\mathrm{K}(R_\mathrm{t}) \, ,
\end{equation}
with the correction factor for the non-Keplerian rotation $K_\mathrm{acc}= 0.775$, $m_\mathrm{acc}=-0.147$, and the Keplerian rotation rate at the truncation radius $\Omega_\mathrm{K}(R_\mathrm{t})$.

\textit{Magnetic star-disk interaction:}
The influence of the magnetic star-disk interaction in \cite{Ireland21} is modeled as magnetospheric ejections \citep[MEs,][]{matt05}.
MEs can transfer angular momentum between the star and the disk due to the twist of magnetic field lines that are rooted at the stellar surface and coupled to the disk material \citep[e.g.,][]{Armitage96, matt05, Zanni13, Ireland21}. 
The amount of angular momentum transferred to or away from the star depends on the relation between the truncation radius $R_\mathrm{t}$ and the co-rotation radius 
\begin{equation}
    R_\mathrm{co} = \left[ K_\mathrm{acc} \left( \frac{R_\mathrm{t}}{R_\star}\right)^{m_\mathrm{acc}} \right]^{2/3} \left( \frac{G M_\star}{\Omega^2} \right)^{1/3} \, ,
\end{equation}
which includes the effect of non-Keplerian rotation.
Disk material, which is coupled to the stellar magnetic field lines inside (outside) $R_\mathrm{co}$ rotates faster (slower) compared to the star and increases (decreases) the stellar angular momentum.
This process is highly dynamic and complex.
Thus, the following formulation should be treated with caution and interpreted as a time-averaged approximation.
Following \cite{Ireland21}, the total torque due to MEs can be written as
\begin{equation}\label{eq:tor_me}
    \Gamma_\mathrm{ME} = K_\mathrm{ME} B_\star^2 R_\star^3 \left( \frac{R_\mathrm{t}}{R_\star} \right)^{m_\mathrm{ME}} \left[ 1 - \left(\frac{R_\mathrm{t}}{R_\mathrm{co}} \right)^{3/2} \right] \, ,
\end{equation}
with numerical fit parameters $K_\mathrm{ME} = 0.00772$ and $m_\mathrm{ME}= -2.54$.
If $R_\mathrm{t} < R_\mathrm{co}$ ($R_\mathrm{t} > R_\mathrm{co}$), the total torque due to MEs is positive (negative) and the star spins up (spins down).

\textit{Accretion-powered stellar wind (APSW):}
In the picture of an accretion-powered stellar wind, a certain fraction $W$ of the accretion rate is ejected by the star; $\Dot{M}_\mathrm{w} = W \Mdot$.
According to \cite{Weber67}, the torque due to such a stellar outflow can be written as 
\begin{equation}\label{eq:apsw}
    \Gamma_\mathrm{w} = - |\Dot{M}_\mathrm{w}| \Omega_\star r_\mathrm{A}^2 \, ,
\end{equation}
with the Alfvén radius $r_\mathrm{A}$. 
The Alfvén radius can be related to the open stellar magnetic flux \citep[][]{Reville2015} and based on numerical best-fit parameters \citep[][]{Ireland21}, the torque due to an APSW can be written as
\begin{equation}\label{eq:gamma_w}
    \Gamma_\mathrm{w} = -K_\mathrm{w1} |\Mdot| (G M_\star R_\star)^{1/2} W^{m_\mathrm{w1}} f^{m_\mathrm{w2}} \left[ 1+ \left( \frac{f}{K_\mathrm{w2}} \right)^2 \right]^{-m_\mathrm{w3}} \left( \frac{R_\mathrm{t}}{R_\star} \right)^{m_\mathrm{w4}} \, ,
\end{equation}
with best-fit parameters $K_\mathrm{w1} = 57.92$, $m_\mathrm{w1} = 0.254$, $m_\mathrm{w2} = 1.092$, $K_\mathrm{w2} = 0.0356$, $m_\mathrm{w3} = 0.373$, $m_\mathrm{w4} = 0.195$, and the fraction of the break-up rotation rate $f = \Omega_\star / \Omega_\mathrm{crit}$. The torque due to stellar winds $\Gamma_\mathrm{w}$ always removes AM from the star.

\subsection{Effects of different stellar rotation rates}\label{sec:model_omega}

Based on this spin evolution model presented above, we can predict the behavior of a star that rotates at a certain rate $\Omega_\mathrm{old}$ and spins up to $\Omega_\mathrm{new} > \Omega_\mathrm{old}$, with all other parameters unchanged. 
First, the accretion torque $\Gamma_\mathrm{acc}$ is reduced (see \equo{eq:gamma_acc}). 
Second, the torque due to the magnetic star-disk interaction $\Gamma_\mathrm{ME}$ is smaller as the co-rotation radius moves closer to the truncation radius for fast rotation rates (see \equo{eq:tor_me}). 
Third, the rate of angular momentum that is removed by the APSW, $\Gamma_\mathrm{w}$, increases (see \equo{eq:apsw}).
An increasing stellar rotation rate ($\Omega_\mathrm{new} > \Omega_\mathrm{old}$) would thus result in a spin-down torque and the star would return to $\Omega_\mathrm{old}$.
Based on the same principles, a decreasing stellar rotation rate ($\Omega_\mathrm{new} < \Omega_\mathrm{old}$) would result in a spin-up torque.
Hence, $\Omega_\mathrm{new}$ evolves back to $\Omega_\mathrm{old}$, and the timescale, on which this change occurs is studied in \sref{sec:parameters}.

\subsection{Accretion rate onto the star}

\begin{figure}
    \centering
         \resizebox{\hsize}{!}{\includegraphics{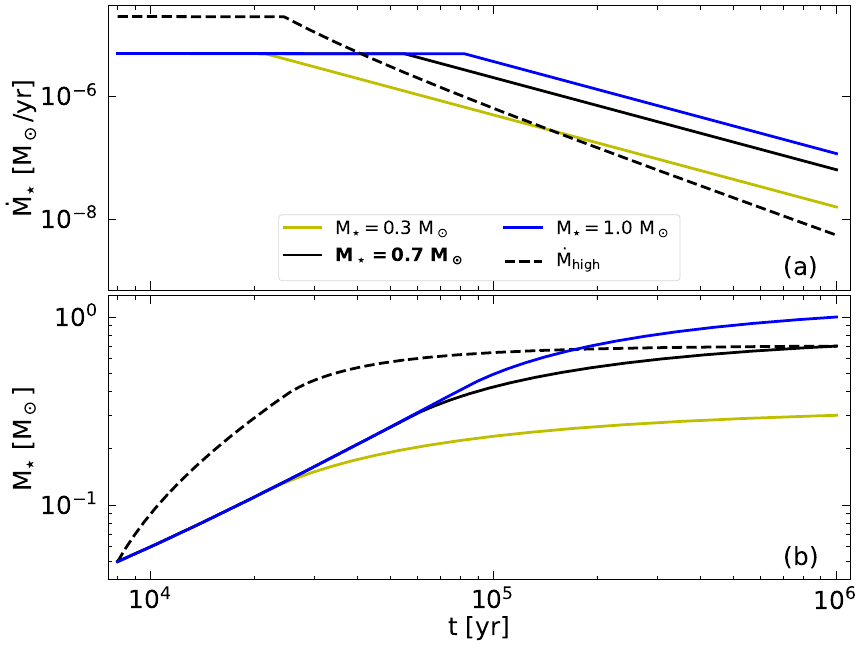}}
    \caption{
    Panel (a) shows the accretion rate and Panel (b) shows the stellar mass over time. The yellow, black, and blue lines represent a final stellar mass of 0.3, 0.7, and 1.0~$\mathrm{M_\odot}$, respectively.
    The black dashed line ($\Dot{M}_\mathrm{high}$) represents a different accretion history, which also results in a final stellar mass of 0.7~$\mathrm{M_\odot}$ at 1~Myr.
    }
    \label{fig:accretion}
\end{figure}

In this study, we start from an initial mass of $M_\mathrm{init} = 0.05$ and consider three different stellar masses at an age of 1~Myr; $M_\mathrm{final} = 0.3,~0.7,~\mathrm{and}~1.0~\mathrm{M_\odot}$.
The accretion rate is defined in two regimes. Initially, we assume a high, constant accretion rate of $\Dot{M}_\mathrm{init} = 5\times10^{-6}~\Msolpyr$, which is equivalent to the mass infall rate $\Dot{M}_\mathrm{infall} \propto c_\mathrm{s}^3 / G$ with the isothermal sound speed $c_\mathrm{s}$ and an external temperature of $\approx 21$~K \citep[e.g.,][]{Vorobyov17c}.
Thus, an initial stellar mass of $M_\mathrm{init} = 0.05$ corresponds to an age of $\sim 10^4$~years, assuming a roughly constant accretion rate during this time.
After a specific time $t_\mathrm{acc}$, the accretion rate decreases according to a power law with an exponent of -1.5, $\Dot{M}_\star(t)\propto t^{-1.5}$.
A stellar mass of 0.3, 0.7, and 1.0~$\mathrm{M_\odot}$ can be obtained at 1~Myr for $t_\mathrm{acc}=21.4$, 54.5 and 81.5~kyr, respectively. 
\fig{fig:accretion} shows the evolution of the stellar accretion rate (Panel a) and the stellar mass (Panel b) for a final stellar mass of 0.3, 0.7 and 1.0~$\mathrm{M_\odot}$ (solid colored lines).
To study the influence of a different accretion history, we increase the initial accretion rate to $\Dot{M}_\mathrm{init} = 2\times10^{-5}~\Msolpyr$ and decrease the power-law exponent to $-2$ ($\Dot{M}_\mathrm{high}$ dashed line).
The final stellar mass at 1~Myr is also 0.7~$\mathrm{M_\odot}$ in case of using $\Dot{M}_\mathrm{high}$.

\subsection{Parameter space}

Finding reasonable ranges for stellar and accretion parameters for very young, embedded objects is challenging due to theoretical and observational uncertainties and a limited number of observations.
The ranges for the free parameters presented below are based on previous studies, theoretical models, and observational results.

\textit{Magnetic field strengths:}
One important stellar parameter that influences the spin evolution model presented above is the magnetic field strength.
It is assumed that the large-scale dipole field components spin evolution of young stars \citep[e.g.,][]{Finnley18}.
Small-scale field components, on the other hand, play only a minor role.
We note that the results presented in \cite{Finnley18} do not cover very young, embedded objects with high accretion rates (see \sref{sec:applicability}).
Due to observational limitations, there is only a limited number of magnetic field estimates for Class~I~(0) objects. 
Based on the observation of magnetically-sensitive line broadening (Zeeman effect), a total magnetic field strength of $2.9\pm0.4$~kG for WL~17 \citep[][]{Krull09} and $1.36\pm0.06$~kG for V347~Aur \citep[][]{Flores2019} can be inferred.
Unfortunately, the total field strength combines the small and large-scale features of the magnetic field and a distinction is only possible with additional (more precise) measurement \citep[e.g., Zeeman-Doppler imaging, ZDI,][]{Johnstone14}.
Such observations are currently not available for embedded Class~I~(0) objects.
In addition, the small number of available measurements does not have to be representative.
There are, however, estimates on the magnetic field strength of Class~I~(0) objects ranging from $\sim 0.1$~kG to $\sim 1$~kG based on flux conservation during the core collapse \citep[][]{Tsukamoto2022}, near-infrared K-band spectra \citep[][]{Laos2021}, and laboratory plasma experiments \citep[][]{Burdonov2021}.
These estimates match the inferred field strengths for WL~17 and V347~Aur. 
Furthermore, \cite{Laos2021} argue that Class~I~(0) objects should have similar accretion mechanisms and magnetic field strengths to Class~II objects.
For these slightly further evolved objects, the large-scale field component ranges from $\sim 0.1$~kG to $\gtrsim 1$~kG \citep[e.g.,][]{Johnstone14}.
Thus, we choose a range for the large-scale magnetic field strength, $B_\star$, from 0.1~kG to 2.0~kG.
These values are kept constant throughout the simulations in \sref{sec:results}. 
In addition, we explore the possibility of mass-dependent magnetic field strengths as presented in \cite{Browning16} (see \sref{sec:slowfast}).

\textit{Initial rotation rates:}
While there are rotational studies of young ($\sim 1$~Myr) T~Tauri stars in NGC~2264 \citep[][]{Venuti2017}, ONC \citep[][]{Rebull2006}, $\sigma$~Ori \citep[][]{Cody2010}, and Taurus \citep[][]{Rebull2020} , there are no large-scale surveys of stellar rotation periods of Class~I~(0) objects.
Similar to magnetic fields, there are only singular estimates for individual stars.
The Class I object Elias~29 has a rotation period of $\sim 2$~days \citep[][]{Pillitteri2019}.
Furthermore, \cite{White2007} suggest that Class~I objects rotate slightly faster than their Class~II counterparts.
This would result in rotation periods of $\sim 1$~day during the Class~I phase.
The initial rotation periods at the beginning of the Class~I~(0) phase, however, are not well constrained.
Thus, we explore a wide range of initial rotation rates from very slow-rotating stars with $\Omega_\mathrm{init} = 0.01 \Omega_\mathrm{crit}$ to fast-rotating stars with $\Omega_\mathrm{init} = 0.9 \Omega_\mathrm{crit}$, with $\Omega_\mathrm{crit}$ being the critical or brake-up rotation rate, which is usually estimated as \citep[e.g.,][]{Maeder2009}
\begin{equation}
    \Omega_\mathrm{crit} = \sqrt{\frac{G M_\star}{(1.5 R_\star)^3}} \, .
\end{equation}

\textit{Initial stellar radius:}
Studies of early stellar evolutionary stages that start shortly after the second collapse and the formation of the second hydrostatic core (with an initial stellar mass of $0.01~\mathrm{M_\odot}$) assume stellar radii of $0.25~\mathrm{R_\odot}$ to $3.0~\mathrm{R_\odot}$ \citep[e.g.,][]{Hosokawa2011, Kunitomo17, Steindl21}.
According to \cite{Kunitomo17}, initial stellar radii of $\gtrsim 1.5~\mathrm{R_\odot}$ are required to explain the most luminous stars and small initial radii can explain the low luminosities, and the observed luminosity spread in young clusters \citep[][]{Hillenbrand2009, Soderblom2014}.
Although we start with slightly further evolved initial stellar models (with an initial mass of $0.05~\mathrm{M_\odot}$), we assume similar initial stellar radii ranging from $0.6~\mathrm{R_\odot}$ to $2.5~\mathrm{R_\odot}$.

\textit{Accretion powered stellar wind efficiency:}
In this study, winds originating from the star are represented by accretion-powered stellar winds (APSW).
Such winds extract a certain amount of angular momentum from the star and cause a spin-down (see \equo{eq:apsw}).
The exact driving mechanisms of these winds are still under debate \citep[see, e.g., the discussion in][]{Gallet19}.
We adapt the driving mechanisms proposed by \cite{Decampli81} and \cite{Matt05APSW}.
Angular momentum can be removed from the star by Alfvén waves, excited by the accreting material at the stellar surface and traveling outwards.
The mass-loss rates of such winds are assumed to be in the order of one percent of the accretion rate onto the star \citep[$W\sim 1$\%, e.g.,][]{Crammer08, Pantolmos20}. 
Thus, the APSW efficiency value $W$ ranges from 0\% to 5\% in this study.
We note that there are also other wind-driving mechanisms affecting the star-disk system (e.g., jets launched at the inner disk).
We discuss the possible impact of such winds or jets on the rotational evolution of the young star in \sref{sec:disk_winds}.

\subsection{Reference model}

We define a reference model, to qualitatively show how each parameter affects the rotational evolution of young stars.
In the following sections, the reference values are marked in boldface. 
The reference parameters are summarized in \tab{tab:ref_mod}.

\begin{table}[ht]
\centering
\caption{Reference model parameters used in this study. We note that the stellar mass corresponds to an age of 1~Myr.}        
\begin{tabular}{c c | c c}         
\hline\hline 
$M_\star$  & $0.7~\mathrm{M_\odot}$ & $\Omega_\mathrm{init}$ & $0.1~\mathrm{\Omega_{crit}}$ \\
$B_\star$ & 1.0~kG & $R_\mathrm{init}$ & $1.1~\mathrm{R_\odot}$ \\
$W$ & 2\%  & $\beta_\mathrm{up}$ & 0.2 \\
$M_\mathrm{add}$ & 0.1 & & \\

\hline\hline                                            
\end{tabular}
\label{tab:ref_mod}  
\end{table}


\section{Results}
\label{sec:results}

\subsection{Evolution of the reference value}

\begin{figure}
    \centering
         \resizebox{\hsize}{!}{\includegraphics{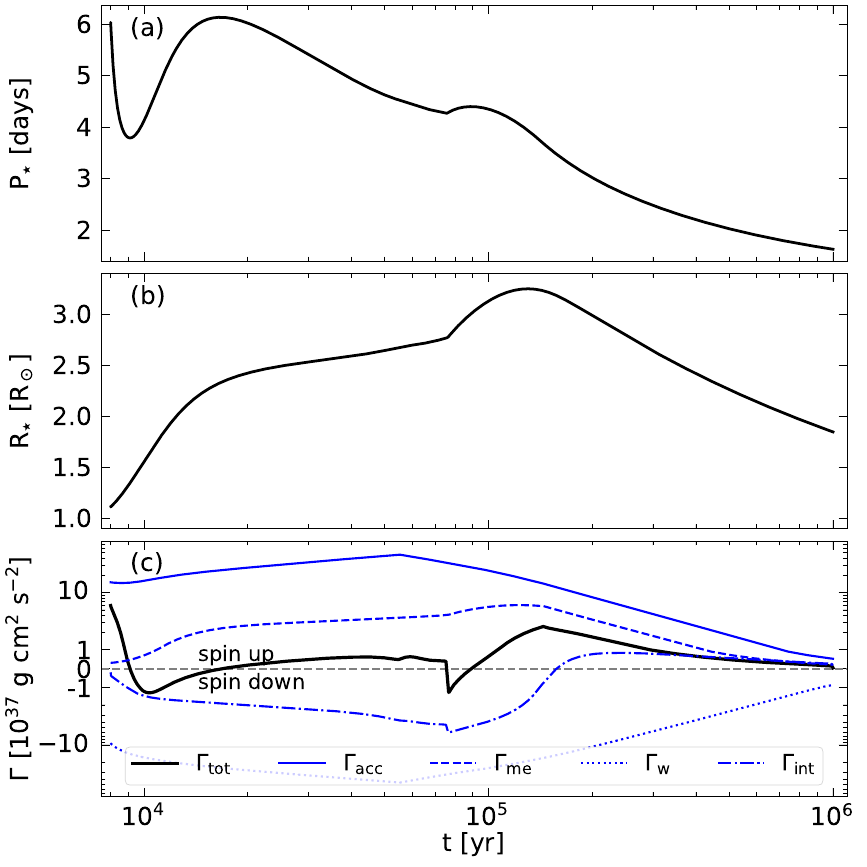}}
    \caption{
    Evolution of the reference model. The stellar rotation period and radius are shown in Panels (a) and (b), respectively.
    The different torque contributions are shown in Panel (c).
    The accretion torque, $\Gamma_\mathrm{acc}$, the influence of the magnetic star-disk interaction, $\Gamma_\mathrm{ME}$, the APSW torque, $\Gamma_\mathrm{W}$, and the internal contributions due to variation of the stellar moment of inertia, $\Gamma_\mathrm{int}$, are shown as a solid blue line, dashed blue line, dotted blue line, and dash-dotted blue line, respectively.
    The total torque, $\Gamma_\mathrm{tot}$, defined as the sum of the other torque contributions is shown as a solid black line.
    A grey dashed line marks the change between the spin-up and spin-down regime.
    }
    \label{fig:ref}
\end{figure}

Before going into detail on the individual parameters, we want to show the evolution of the reference model (see \fig{fig:ref}).
The choice of initial conditions leads to an initial rotation period of 6~days.
During the first few thousand years, the star spins up to $\approx 4$~days. 
This initial spin-up can be explained by adjusting the rotation period to the accretion rate when starting the simulation.
Starting from the initial stellar radius of $R_\mathrm{init} = 1.1~\mathrm{R_\odot}$, the star inflates due to the energy of the accreting material added into the stellar interior and the onset of Deuterium burning. 
Up to an age of $\approx 18$~kyr, the inflation of the star (combined with the APSW) can counteract the spin-up influence of accretion and the star-disk interaction and the star spins down towards 6~days ($\Gamma_\mathrm{tot}<0$, in Panel c of \fig{fig:ref}).
With the increasing stellar radius, the spin-up influence of the MEs increases as well, and the stellar period decreases toward 4~days up to an age of 80~kyr.
At $\sim 80$~kyr, the stellar inflation increases for a short period of time, resulting in a brief spin-down.
Afterward, the star contracts for the remaining simulation, reaching a final stellar radius of $1.84~\mathrm{R_\odot}$ after 1~Myr.
During contraction, $\Gamma_\mathrm{int}$ increases the stellar spin up and the star reaches a period of 1.6~days after 1~Myr.

\subsection{Effects of individual parameters}\label{sec:parameters}

\begin{figure*}
    \centering
         \resizebox{\hsize}{!}{\includegraphics{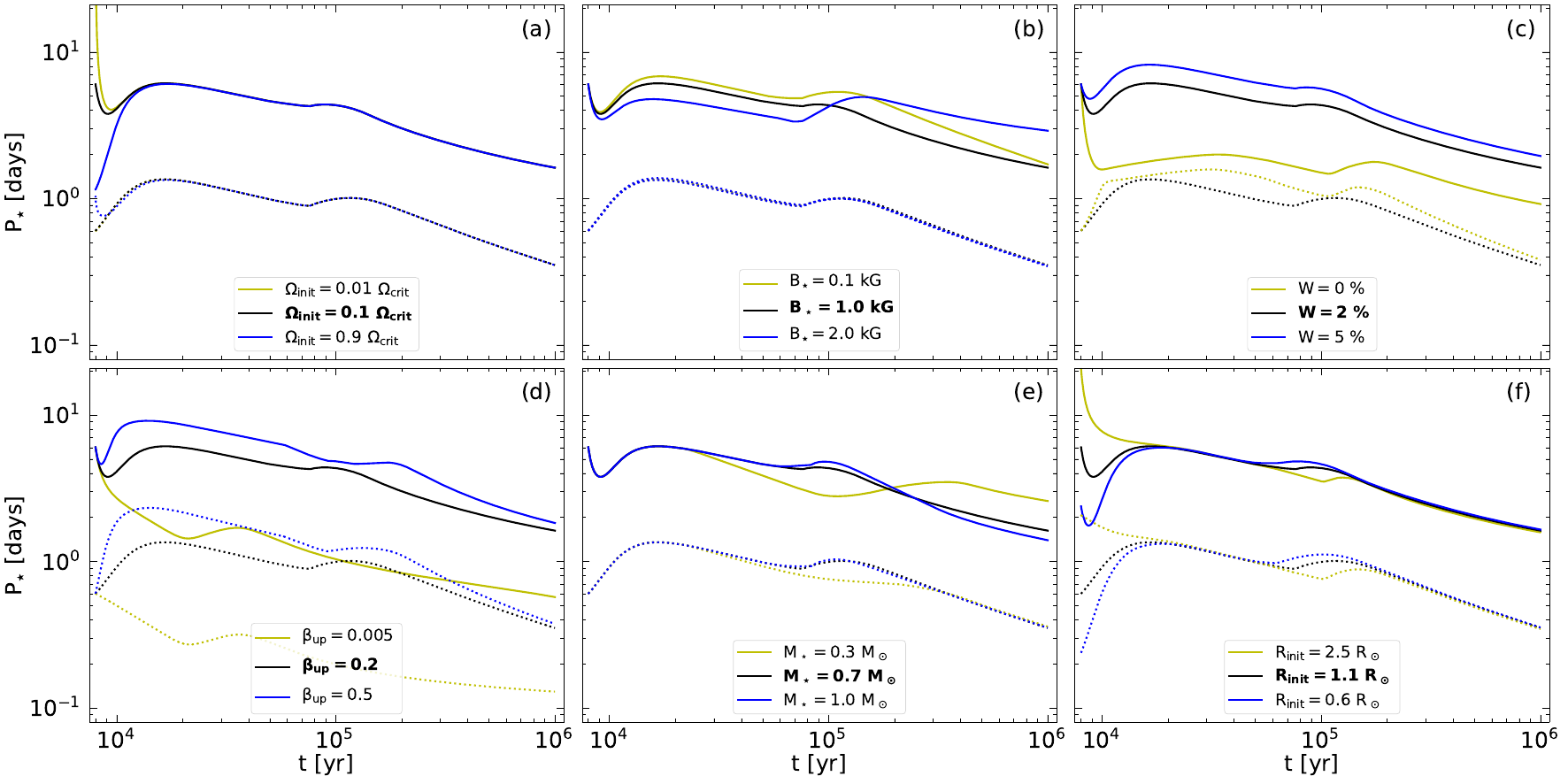}}
    \caption{
    Evolution of the stellar rotation period, $P_\star$, for each parameter variation:
    the initial rotation rate $\Omega_\mathrm{init}$ (Panel a), the stellar magnetic field strength $B_\star$ (Panel b), the APSW efficiency $W$ (Panel c), the amount of energy of the infalling material added into the star, $\beta_\mathrm{up}$ (Panel d), the final stellar mass at 1~Myr, $M_\star$ (Panel e), and the initial stellar radius, $R_\mathrm{init}$ (Panel f).
    The reference values are marked in boldface.
    For comparison, the critical rotation period is shown for each evolution as dotted lines.
    }
    \label{fig:rot}
\end{figure*}

\begin{figure*}
    \centering
         \resizebox{\hsize}{!}{\includegraphics{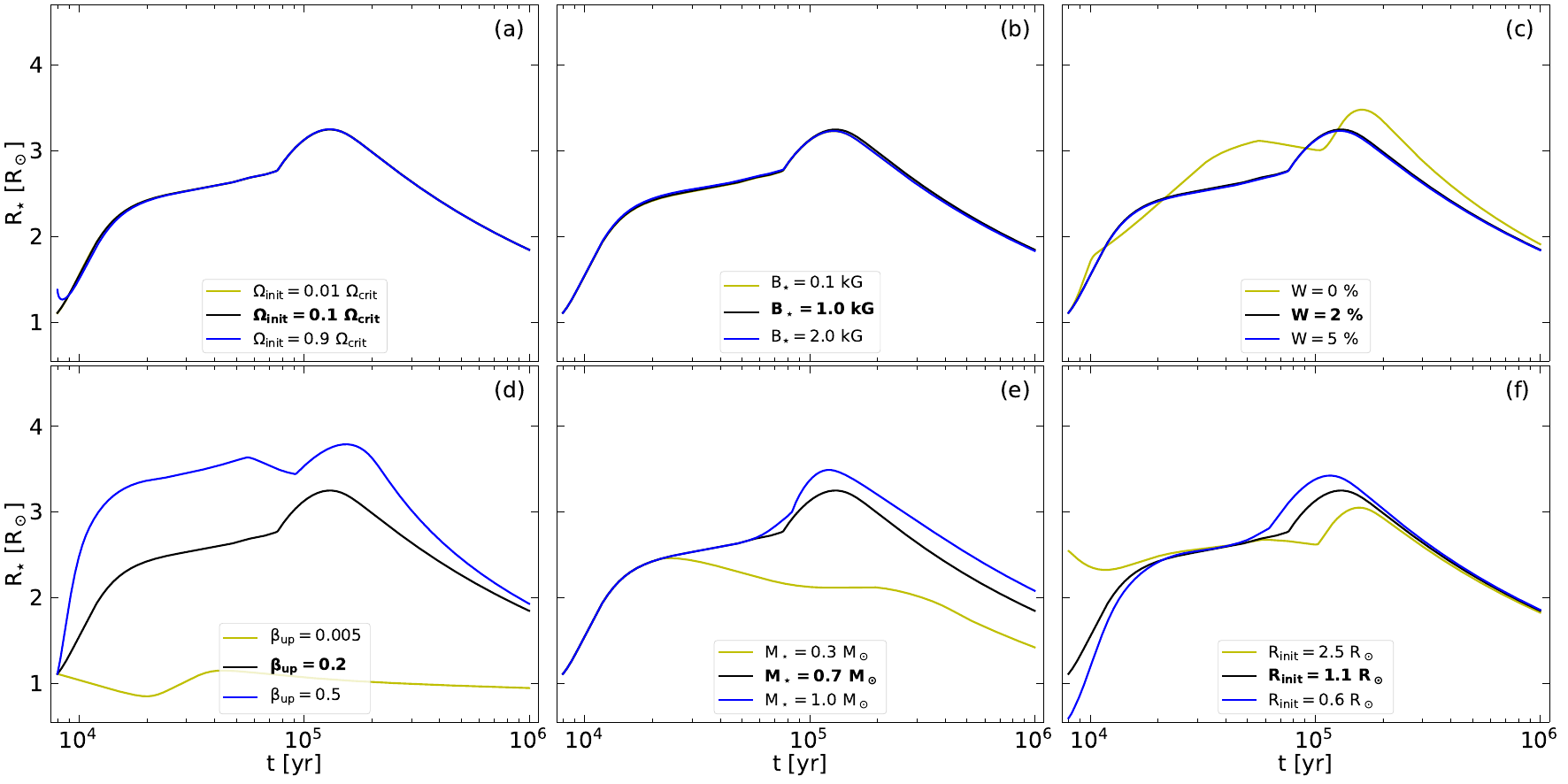}}
    \caption{
    Evolution of the stellar radius, $R_\star$, for each parameter variation (same as in \fig{fig:rot}).
    The reference values are marked in boldface.
    }
    \label{fig:radii}
\end{figure*}

Now we examine how the stellar rotational evolution is affected by the initial rotation rate $\Omega_\mathrm{init}$, the stellar magnetic field strength $B_\star$, the APSW efficiency $W$, the initial stellar radius $R_\mathrm{init}$, the final stellar mass at 1~Myr $M_\star$, and the amount of energy of the infalling material added into the star, $\beta_\mathrm{up}$ (see \fig{fig:rot}). For a clear description of different effects, we show, in addition to the stellar rotation period, the evolution of the stellar radius in \fig{fig:radii}, the individual torque contributions in \app{sec:tor_comp}, and the HRD for each model in \fig{fig:hrd}.

\textit{The initial rotation rate:}
First, we want to show the effects of different initial rotation rates, $\Omega_\mathrm{init}$ (Panel a in \fig{fig:rot} and \fig{fig:radii}). 
We vary $\Omega_\mathrm{init}$ from 1\% to 90\% of the critical (or break-up) velocity.
The black, blue, and yellow lines symbolize the reference, fast, and slow rotating models.
Within the first $\sim 5$~kyr, the initial difference in the stellar periods of $\approx 59$~days converges towards a common value (see \sref{sec:model_omega}).
As a consequence of the fast initial rotation and strong centrifugal forces, the initial stellar radius for a star with $\Omega_\mathrm{init} = 0.9~\mathrm{\Omega_{crit}}$ is slightly larger compared to the other models.
However, this difference disappears after a few thousand years and the stellar evolution toward 1~Myr is completely unaffected by the initial rotation rate $\Omega_\mathrm{init}$.

\textit{Stellar magnetic field strength:}
The stellar magnetic field strength does affect all external torque contributions, such as $\Gamma_\mathrm{acc}$, $\Gamma_\mathrm{ME}$, and $\Gamma_\mathrm{W}$.
The torque due to MEs, however, is affected most strongly (see \equo{eq:tor_me} in \sref{sec:spin_model}).
As shown in Panel (b) of \fig{fig:rot}, a strong magnetic field causes the star to spin up during the first $\sim 100$~kyr.
After $\sim 100$~kyr, a strong magnetic field causes a spin down.
The strong and weak magnetic field strengths are symbolized by the blue and yellow lines, respectively.
This behavior can be explained with \equ{eq:tor_me}.
During the initial $\sim 100$~kyr, the star is still heavily embedded and the accretion rates can be assumed to be high (see \fig{fig:accretion}).
The truncation radius, $R_\mathrm{t}$, is pushed inwards towards the star (regardless of the magnetic field strengths used in this study), and the ratio between $R_\mathrm{t}$ and the co-rotation radius, $R_\mathrm{co}$, is small.
Thus, the torque due to the magnetic star-disk interaction, $\Gamma_\mathrm{ME}$, is positive, and a stronger magnetic field strength results in a stronger spin-up torque.
After $\sim 100$~kyr, the accretion rate decreases, and a stronger magnetic field pushes the truncation radius outwards closer to the co-rotation radius.
The disk's rotation at the truncation radius, $\Omega(R_\mathrm{t})$, and the difference between $\Omega(R_\mathrm{t}) - \Omega_\star$ decreases.
Consequently, stronger magnetic fields result in a weaker accretion spin-up torque (see \equo{eq:gamma_acc}), a weaker torque due to the magnetic star-disk interaction, $\Gamma_\mathrm{ME}$ (see \equo{eq:tor_me}), and a slower rotation period compared to weak magnetic field strengths.
The differences in the rotation periods are relatively small ($\sim 1$~day) and the stellar radius evolution is almost unaffected (see Penal b of \fig{fig:radii}).

\textit{APSW efficiency $W$:}
In the current spin evolution model, a stellar wind removes angular momentum from the star and causes a spin-down.
The amount of AM that is removed scales with the mass loss, which is proportional to the APSW efficiency $W$ (see \equo{eq:apsw}).
Thus, a higher efficiency value results in a stronger spin-down torque, and the star slows down compared to a weaker or no APSW (see Panel d in \fig{fig:rot}).
The difference, especially during the early evolution ($\lesssim 10^5$~years) can be significant.
With no angular momentum removed by an APSW ($W=0$\%), the star spins up to $81.7$\% of the critical rotation rate (see the dotted lines in \fig{fig:rot}).
The stellar rotation period differs by more than 7 days, equivalent to a factor of $\approx 4$.
The stellar radius is also affected by the different rotation
periods.
A fast-rotating star shows a larger radius (Panel c in \fig{fig:radii}) and a lower effective temperature (see Panel c in \fig{fig:hrd}).
Thus, the first nuclear reactions ignite at a later age (see the shifted radius evolution in Panel c of \fig{fig:radii}).
While, toward an age of 1~Myr, the rotation periods still show a difference of a factor of $\sim 2$, the stellar radii converge to a similar value.

\textit{Maximum amount of additional heat $\Dot{E}_\mathrm{add}$:}
The amount of additional heat can be controlled by the free parameter $\beta_\mathrm{up}$\footnote{We expect that the stellar evolution is largely independent of $\beta_\mathrm{low}$, as long the value is chosen to be small enough, $\beta_\mathrm{low} \ll 0.1$.}.
The value of $\beta_\mathrm{up}$ does have a significant impact on stellar evolution.
While the evolution of most stars can be explained by a substantial amount of additional heat added to the star \citep[ corresponding to $\beta_\mathrm{up}\gtrsim 0.1$,e.g.,][]{Vorobyov17, Kunitomo17, Steindl21}, objects with very low luminosity require $\beta \sim \beta_\mathrm{low}$ \citep[][]{Kunitomo17, Vorobyov2017VELLOs}. 
The amount of additional heat affects the evolution of the stellar radius.
The more energy added to the stellar interior, the stronger the star inflates (see Panel d in \fig{fig:radii}).
In the cold-accretion scenario ($\beta=\beta_\mathrm{low}$), the stellar radius does not inflate and remains approximately constant at a value of $\sim 1~\mathrm{R_\odot}$.
These different evolution tracks influence stellar rotation (see Panel d in \fig{fig:rot}).
The strong initial inflation to $R_\star > 3~\mathrm{R_\odot}$ for $\beta_\mathrm{up} = 0.5$ results in a spin-down to 9.2~days.
With larger radii, MEs become more effective and start to spin up the star slowly.
After 1~Myr, the stellar period has reached $\sim 2$~days.
Without the initial inflation, the star continuously spins up during its evolution (excluding the short period of inflation due to the first nuclear reactions).
After 1~Myr, the stellar period has reached $\sim 0.57$~days.
Within the parameter ranges studied so far, a variation in $\beta_\mathrm{up}$ marks the largest impact on the stellar rotation period.

\textit{Stellar mass at 1~Myr:}
To reach different stellar masses at 1~Myr, we vary the age, at which the initially constant accretion rate decreases (see \fig{fig:accretion}).
During the first $\sim 10^5$~years, smaller final stellar masses result in faster-rotating stars (see Panel e in \fig{fig:rot}).
To explain this behavior, the amount of additional heat has to be considered.
For a decreasing accretion rate, $\beta_\mathrm{up}$ and consequently $\Dot{E}_\mathrm{add}$ decrease as well.
Thus, the stellar inflation stops and the star starts contracting, resulting in a stellar spin-up.
After $10^5$~years, the influence of the additional heat diminishes as accretion rates decrease further. 
Now, a lower accretion rate results in less accretion torque, $\Gamma_\mathrm{acc}$, and a smaller stellar radius results in less effective MEs.
As a consequence, a smaller accretion rate causes a slower rotation compared to high accretion rates.

\textit{Initial stellar radii:}
Finally, we show the impact of different initial stellar radii on the spin evolution.
For a given stellar mass, a larger radius results in a cooler temperature throughout the star. 
Starting at a large initial stellar radius, gravity prevails over the inflation due to the additional heat, and the star contracts slightly (see Panel f in \fig{fig:radii}).
After $\sim 20$~kyr, the inflation of the stellar radius has balanced the initial differences.
The temperature deep in the stellar interior, however, is still cooler in case of an initially larger radius and the first nuclear reactions start slightly later (maximum radii are reached later).
Reaching 1~Myr, the stellar radii converge toward a common value and the initial differences are erased.
A similar picture emerges for the stellar rotation periods.
Initial differences converge fast and throughout the evolution, only small differences arise due to different stellar radii.
After 1~Myr, the rotation period evolved independently of the initial stellar radius.
In our reference case, we choose $\beta_\mathrm{up} = 0.2$, which corresponds to a warm accretion scenario.
\cite{Kunitomo17}, however, pointed out that the importance of the initial stellar radius increases, when decreasing the amount of additional heat.
Hence, additional simulations have been carried out with different initial radii and $\beta_\mathrm{up} = 0.005$ (see \app{sec:rin_beta}).
Without the additional heat and the resulting stellar inflation, the stellar radii maintain a significant difference throughout their evolution and after 1~Myr, the stellar radii differ by a factor of $\sim 2$ (see Panel b in \fig{fig:rin_beta}).
As a result, the stellar rotation period does not converge and shows a dependence on the initial stellar radius.
Smaller stellar radii result in faster-rotating stars and after 1~Myr, the rotation periods differ by a factor of 2.6 (see Panel a in \fig{fig:rin_beta}).

\subsection{Other factors affecting $\Dot{E}_\mathrm{add}$}\label{sec:change_eadd}

The additional heat, $\Dot{E}_\mathrm{add}$, is not only controlled by $\beta$.
The region, in which $\Dot{E}_\mathrm{add}$ is added, and the accretion history can influence the stellar evolution.
We reduce the region, in which the additional heat, $\Dot{E}_\mathrm{add}$, is distributed to $M_\mathrm{add} = 0.01$.
With more additional heat injected close to the stellar surface, the initial stellar inflation increases (see the sharp increase of the stellar radius in Column a of \fig{fig:discussion}).
During the whole evolution, the stellar radius remains slightly larger compared to the reference model.
Reaching 1~Myr, the stellar radius converges toward the reference value as the influence of $\Dot{E}_\mathrm{add}$ becomes less important as the accretion rate decreases.
As discussed before, larger inflated stars tend to rotate more slowly.
During the strong initial inflation, the difference to the reference model is the largest ($\sim 3$~days).
Similar to the stellar radius, reaching an age of 1~Myr, the stellar period converges toward the reference value.

Next, we change the accretion history.
Starting from an initial accretion rate of $2\times10^5~\Msolpyr$, which is factor 4 larger compared to the reference value, the accretion rate drops earlier and faster to reach the same stellar mass ($0.7~\Msol$) compared to the reference model (see \fig{fig:accretion}).
The strong accretion rate injects a large amount of additional heat into the stellar interior and the star inflates rapidly to $> 4~\mathrm{R_\odot}$ (see Column a of \fig{fig:discussion}).
Due to this immense expansion, the star slows down to over 10~days. 
After the accretion rate drops, the stellar expansion stops and the star contracts toward the reference value, reaching the end of the simulation.
In contrast to the previous model with $M_\mathrm{add} = 0.01$, the smaller accretion rate at ages $>10^5$~years reduces the accretion torque and the torque due to MEs and the star spins down.
At the age of 1~Myr, the stellar rotation period differs by a factor of 1.9.

While the region, in which the additional heat is injected, plays only a minor role in the stellar spin evolution, the accretion history significantly affects the stellar period.
In our model, the accretion rate is assumed to be smooth, without short-term variations.
This idealistic assumption, however, should merely be considered as an average over time.
In hydro-dynamic disk simulations, especially the embedded phase ($<10^5$~years), show episodic outbursts with peak accretion rates of $>10^5~\Msolpyr$ \citep[e.g.,][]{Vorobyov15, Vorobyov17c, Hsieh2019, Kadam2021}.
During such an outburst, a huge amount of additional heat is injected into the stellar interior, the star inflates, and the rotation period increases.
The nature of these outbursts depends, amongst other things, on the stellar mass.
More massive stars undergo more and stronger outbursts over a larger period of their early life.
If each outburst spins down the star slightly, more massive stars should tend to rotate more slowly compared to low-mass stars.
This trend is, in fact, reported in \cite{Henderson12} for young ($\lesssim 5$~Myr) clusters.

\subsection{Disk winds}\label{sec:disk_winds}

In the presence of a large-scale magnetic field in the accretion disk, disk winds, launched by magneto-rotational processes in the inner disk regions, can remove a significant amount of angular momentum from the disk \citep[e.g.,][]{Zhang2017}.
The exact regions, in which the wind is launched, and the total amount of mass and angular momentum transported away, depend on a variety of factors \citep[e.g.,][Steiner et al. in prep]{Guilet2014}.
A detailed description of this process would go beyond the scope of this work.
However, we can approximate the effect of a strong disk wind on the stellar rotation.
The loss of angular momentum in the inner disk region causes the disk material to slow down and reduces the accretion torque.
For our simulation, we extend \equo{eq:non_kep} by a factor of $K_\mathrm{DW} = 0.5$.
We note that this factor is most likely not constant and depend on different parameters, but for showing the qualitative effect of a disk wind, this assumption should be viable.
The rotational evolution including a disk wind is shown in Column (b) of \fig{fig:discussion} (yellow line).
As expected, the reduced accretion torque leads to a rapid spin-down during the radius inflation.
Reaching 1~Myr, the rotational period is still slower by a factor of 2.0 compared to the reference model, which is similar to the effect of the model with $\Dot{M}_\mathrm{high}$.

\subsection{Stellar metallicity}

\begin{figure*}
    \centering
         \resizebox{\hsize}{!}{\includegraphics{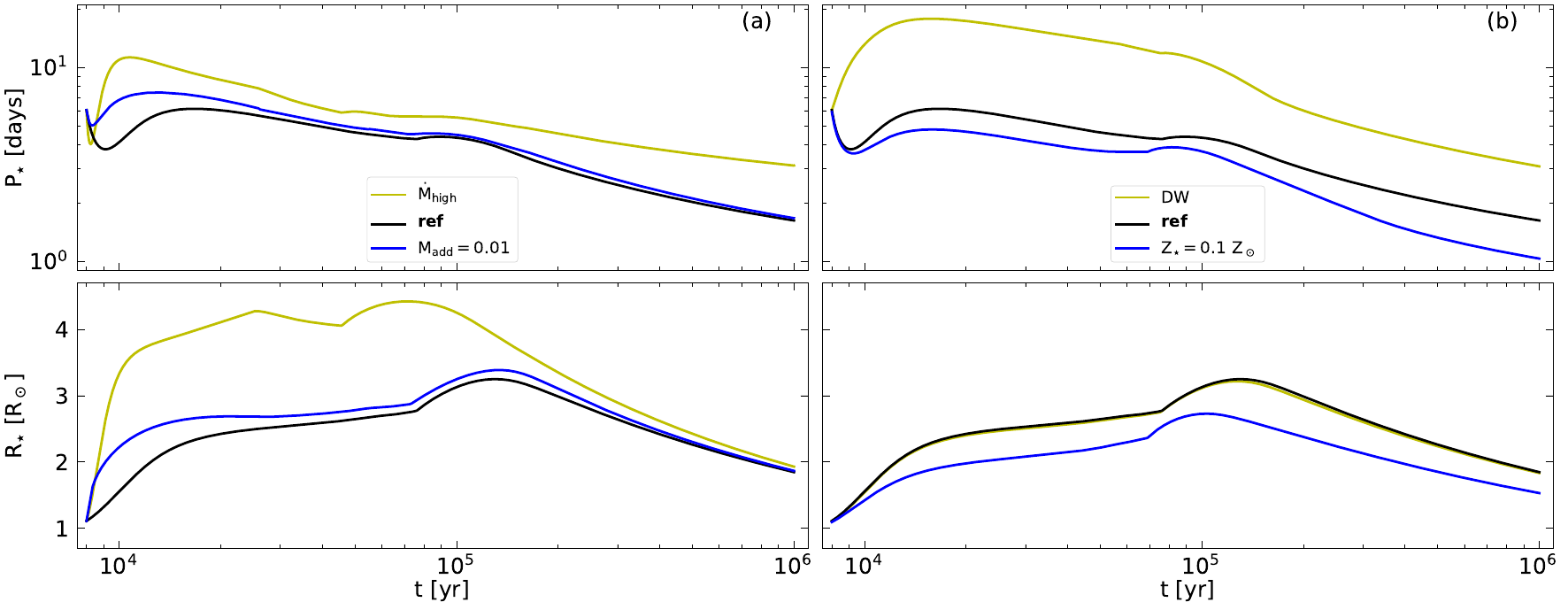}}
    \caption{
    Evolution of the stellar rotation period, $P_\star$, and the stellar radius, $R_\star$, for different variations from the reference model.
    Column (a): region, in which $\Dot{E}_\mathrm{add}$ is distributed, $M_\mathrm{add} = 0.01$ (blue line) and using a different accretion history, $\Dot{M}_\mathrm{high}$ (see \fig{fig:accretion}, yellow line).
    Column (b): low metallicity, $Z_\star = 0.1~\mathrm{Z_\odot}$ (blue line), and including the influence of disk winds (model DW, yellow line).
    The reference model is marked in boldface.
    }
    \label{fig:discussion}
\end{figure*}

Finally, stellar metallicity can affect the rotational evolution of the star during the pre-main sequence and the main sequence.
Low-metallicity stars are expected to rotate faster compared to their solar-metallicity counterparts \citep[e.g.,][]{Amard19, Amard20, Amard20b, Gehrig2023}. 
As low-metallicity stars are more compact compared to solar metallicities, mechanisms that remove angular momentum from the star and scale with the stellar radius are less effective in a low-metallicity environment during the Class~II phase and beyond the main sequence \citep[e.g.,][]{Amard20, Gehrig2023}.
In addition, the disk lifetime of low-metallicity stars is shorter compared to solar metallicities, and the pre-main sequence spin-up due to contraction starts at an earlier age \citep[][]{Yasui16, Yasui21, Guarcello21, Gehrig2023}.
We find similar results for the evolution during the first million years for a metallicity of $Z_\star = 0.1~\mathrm{Z_\odot}$ (see Column b in \fig{fig:discussion}, blue line).
The initial inflation is weaker and the star is more compact during the whole simulation time compared to the reference model.
Thus, the star spins up during the embedded phase.
Reaching the age of 1~Myr, the APSW is less effective in removing angular momentum from the star due to the smaller stellar radius and the star continues to spin up compared to the reference model.

\subsection{Summarizing the effects on $P_\star$}\label{sec:results_summary}

In this work, we have studied the influence of various stellar and disk parameters on the stellar spin evolution during the first million years of stellar evolution. 
We want to summarize our results in an illustrative way in \tab{tab:final}.
We divide the stellar evolution into two phases: the embedded phase $\lesssim 10^5$~year and the later ($\gtrsim 10^5$~years) phase, which corresponds to the Class~II or T~Tauri phase.
For each parameter variation, we indicate if the star rotates faster (red), slower (blue), or approximately unchanged (grey) compared to the reference model.

\begin{table}[ht]
\centering
\caption{Summary of the influence of different stellar and disk parameters on the stellar spin evolution. For each parameter, we distinguish between 3 different effects: the star rotates faster (red), slower (blue), or approximately unchanged (grey) compared to the reference model.}        
\begin{tabular}{|c|c|c|}         
\hline 
& \multicolumn{2}{c|}{Evolution phase} \\ \cline{2-3}
& Embedded & Class~II \\ 
\hline 
fast $\Omega_\mathrm{init}$ & \cellcolor{gray!25}unchanged & \cellcolor{gray!25}unchanged \\
\hline 
strong $B_\star$ & \cellcolor{red!25}fast & \cellcolor{blue!25}slow \\
\hline 
high $W$ & \cellcolor{blue!25}slow & \cellcolor{blue!25}slow \\
\hline
high $\beta_\mathrm{up}$ & \cellcolor{blue!25}slow & \cellcolor{blue!25}slow \\
\hline
low $M_\star$ at 1~Myr & \cellcolor{red!25}fast & \cellcolor{blue!25}slow \\
\hline
high $R_\mathrm{init}$ ($\beta_\mathrm{up}=0.2$) & \cellcolor{gray!25}unchanged & \cellcolor{gray!25}unchanged \\
\hline 
high $R_\mathrm{init}$ ($\beta_\mathrm{up}=0.005$) & \cellcolor{blue!25}slow & \cellcolor{blue!25}slow \\
\hline
high $\Dot{M}_\star$ & \cellcolor{blue!25}slow & \cellcolor{red!25}fast \\
\hline
$M_\mathrm{add}=0.01$ & \cellcolor{blue!25}slow &\cellcolor{gray!25}unchanged \\
\hline
disk winds (DW) & \cellcolor{blue!25}slow & \cellcolor{blue!25}slow \\
\hline
low $Z_\star$ & \cellcolor{red!25}fast & \cellcolor{red!25}fast \\

\hline                                            
\end{tabular}
\label{tab:final}  
\end{table}



\section{Discussion}
\label{sec:discussion}

\subsection{Applicability of the spin model for embedded stars}\label{sec:applicability}

In this study, we use the stellar spin evolution model presented in \cite{Ireland21} that depends strongly on best-fit parameters and proportionality constants (see \equos{eq:gamma_acc}{eq:gamma_w}).
These parameters are based on models with smaller accretion rates $\Dot{M}_\mathrm{acc} \lesssim 10^{-8}~\Msolpyr$.
Such accretion rates are typically found in older, further evolved systems with ages of $\sim 1$~Myr, which marks the end of our simulations.
In their work, \cite{Ireland21} recommend the use of their formulation for $R_\star \leq R_\mathrm{t} \leq R_\mathrm{co}$, which is the case for all our models.
However, the large accretion rates can result in effects not considered by \cite{Ireland21}.

\textit{Disk truncation radius:}
The disk truncation radius, $R_\mathrm{t}$, depends on the balance between stellar magnetic pressure and ram pressure of the infalling material \citep[e.g.,][]{hartmann16}.
For accretion rates of $\Dot{M}_\mathrm{acc} \lesssim 10^{-8}~\Msolpyr$, the truncation radius is usually located at several stellar radii and the dipolar magnetic field component dominates the magnetic field strength at this position \citep[e.g.,][]{Finnley18, Ireland21}.
During phases of higher accretion rates, however, the truncation radius is pushed toward the star and can reach $R_\mathrm{t} = R_\star$.
Close to the stellar surface, the importance of high-order magnetic field components increases.
Quadro- or octopole components can have field strengths of several kilogauss \citep[e.g.,][]{Kochukhov2020hidden}, which could also affect the position of the truncation radius close to the star.
On the other hand, trends in the evolution of the stellar magnetic field suggest that young, fully-convective stars have axisymmetric fields dominated by a dipole component \citep[e.g.,][]{Folsom2016}.
Thus, we assume that the calculation of the truncation radius as presented in \cite{Ireland21} provides a good approximation for young, fully-convective stars.

\textit{Accretion torque:}
The accretion torque, $\Gamma_\mathrm{acc}$, depends on the 'Keplerianity' of the disk material at the truncation radius.
Angular momentum removed by MEs or stellar magnetic torques results in sub-Keplerian rotation of the disk material \citep[e.g.,][]{bessolaz08, matt12, Ireland21}.
In the current formulation, these effects depend on the proportionality constant $K_\mathrm{acc}=0.775$ and the position of the truncation radius (see \equo{eq:non_kep}).
While the formulation should provide reasonable estimates for the accretion torque for $R_\mathrm{t} > R_\star$, $\Gamma_\mathrm{acc}$ could be underestimated in case the truncation radius would be located inside the stellar radius, due to strong accretion rates or, for example, weak magnetic field strengths.
In such cases, we set $R_\mathrm{t} = R_\star$.
As a result, the ram pressure of the accreting material exceeds the stellar magnetic pressure, and the accreting material could be close to Keplerian rotation.

\textit{Magnetospheric ejections:}
The torque due to magnetospheric ejections, $\Gamma_\mathrm{ME}$, depends, similarly to the truncation radius, on the stellar magnetic field strength and the location of the truncation radius (see \equo{eq:tor_me}).
For truncation radii close to the stellar surface, small-scale magnetic field components could affect the magnitude of $\Gamma_\mathrm{ME}$.
Assuming primarily axisymmetric fields dominated by a dipole component \citep[e.g.,][]{Folsom2016}, the current formulation should provide reasonable estimates for $\Gamma_\mathrm{ME}$.

\textit{APSW torque:}
In the current model, accretion-powered stellar winds (APSW) remove angular momentum from the star \citep[see \equo{eq:gamma_w} and, e.g.,][]{Matt05APSW, Gallet19, Ireland21}.
These studies, however, do not consider the very early, embedded stages in their formulations.
It is unclear how winds emerging from the stellar surface interact with an envelope falling onto the disk and how much angular momentum can be transported away from the star under these circumstances.
In case the envelope around the young star has little or no influence on the APSW, the current formulation with efficiency rates of $W\sim 2\%$ provides a realistic estimate for the torque due to APSW.
If, on the other hand, the infalling envelope interferes with the APSW, the efficiency rate could be reduced significantly, which is demonstrated in this study by $W=0\%$.

In summary, the model presented in \cite{Ireland21} should provide reasonable estimates for the torques acting on the young protostar during the embedded phase with high accretion rates.
The individual relations between stellar and disk parameters are assumed to be valid.
We note, however, that the proportionality constants used in \equos{eq:gamma_acc}{eq:gamma_w} could change when repeating the calculations presented in \cite{Ireland21} with higher accretion rates.

\subsection{Magnitude and distribution of $\beta_\mathrm{up}$}\label{sec:eadd}

The evolution of young stars strongly depends on the amount of additional heat, $\Dot{E}_\mathrm{add}$, added into the stellar interior \citep[e.g.,][]{Baraffe2012, Vorobyov2017VELLOs, Vorobyov17c, Kunitomo17}. 
Usually, $\Dot{E}_\mathrm{add}$ is assumed to be a certain fraction, $\beta$, of the energy of the accreted material.
In the hybrid accretion scenario \citep[e.g.,][]{Vorobyov17c}, $\beta$ depends on the accretion rate on the star, ranging from small values, $\beta_\mathrm{low}$, during phases of low accretion rates to high values, $\beta_\mathrm{up}$, during phases of high accretion rates \citep[e.g.,][]{Jensen2018, Elbakyan2019, Steindl21}.
According to \cite{Kunitomo17}, relatively large values of $\beta\gtrsim0.1$ are necessary to explain the evolution of a majority of stars in young clusters\footnote{We note that for a higher Deuterium abundance lower values of $\beta$ suffice to explain the evolution of a majority of stars \citep[][]{Kunitomo17}. Since a variation in the Deuterium abundance has similar effects on the stellar evolution compared to a variation in $\beta_\mathrm{up}$, we forego the simulation of specific models.}.
Low values of $\beta$ (e.g., $\beta=\beta_\mathrm{low}$), on the other hand, result in stellar luminosities of $\lesssim 0.1~\mathrm{L_\odot}$ and can explain very low-luminosity objects \citep[VeLLOs, e.g.,][]{Vorobyov2017VELLOs}.
To explain the luminosity spread in a young cluster, a distribution of $\beta_\mathrm{up}$ has to be assumed.
The exact processes that constrain the magnitude of $\beta_\mathrm{up}$ are still unknown and more detailed modeling of the accretion process and the distribution of the additional heat is required to formulate reasonable limits. 
Our results show that the amount of additional heat affects the stellar spin evolution significantly.
Low values of $\beta_\mathrm{up}$ result in very fast rotating stars (see Panel d in \fig{fig:rot}).
It would be interesting to test if low-luminosity stars tend to rotate faster compared to objects with higher luminosities.
Unfortunately, the available observational data do not allow any conclusion to be drawn so far.

\subsection{External influences}

The previous results are valid for a single star without any external or environmental influences.
Recently, \cite{Roquette21} studied the influence of external photo-evaporation caused by massive stars in young regions.
The presence of a high-mass star can remove mass from the disk due to high-energy radiation and reduce the disk's lifetime.
The timescale, on which the star and the disk can exchange angular momentum is reduced and the stars start to spin up due to contraction at an early age.
Thus, such stars show faster rotation periods after several million years.
During the initial stellar evolution, however, \cite{Roquette21} assume disk-locking, meaning that the stellar rotation period is kept constant as long as the accretion disk is present.
The effect of external photo-evaporation on the stellar spin evolution during the first million years can be summarized as follows:
During the embedded phase the disk material is effectively shielded by the envelope and external photo-evaporation is assumed to have little effect on the disk evolution \citep[e.g.,][]{Winter2020}.
Without the envelope, on the other hand, the high-energy radiation removes material from the disk, and the disk accretion rate decreases for $\gtrsim 10^5$~years.
As shown in the previous models with reduced accretion rates at $\gtrsim 10^5$~years (e.g., model $\Dot{M}_\mathrm{high}$, see \sref{sec:eadd}), the star spins down. 
While assuming disk-locking, this effect is not seen in \cite{Roquette21}.
To model the extent of the spin-down due to lower accretion rates and the spin-up due to shorter disk lifetimes in a high-energy radiation environment, a combination of our early spin evolution model and a hydrodynamic disk model including the effects of external photo-evaporation will provide valuable insights in the early spin evolution.

Another environmental factor is the tidal interaction between different stars in close encounters, binary, or multiple systems.
Tidal interactions in close encounters can remove disk material and reduce the disk accretion rate and lifetime \citep[e.g.,][]{Winter2018}.
In contrast to external photo-evaporation, these close encounters can act on the star-disk system during all phases of its evolution.
The amount of mass removed, depends on the mass of the ’intruding object’ \citep[e.g.,][]{Vorobyov2020Intruder} and the number of encounters \citep[e.g.,][]{Winter2018}.
After the embedded phase, a reduced accretion rate results in a stellar spin-down (similar to the effect of external photo-evaporation).
During the embedded phase, a reduced accretion rate will reduce $\Dot{E}_\mathrm{add}$ and the stellar inflation.
As a consequence, the star rotates faster (see our model with $M_\star = 0.3~\Msol$, Panel e in \fig{fig:rot}).
On the other hand, close encounters can ignite outburst events \citep[e.g.,][]{Vorobyov2021Encounter} that result in a stellar spin-down.
Which of the two mechanisms predominates is unclear and requires a combined model.

Binaries or multiple stellar companions can have different effects on the disk's lifetime and thus, on the stellar spin evolution.
Accretion disks, truncated by a wide binary companion, are assumed to have shorter lifetimes as mass is removed from the accretion disk \citep[e.g.,][]{Kraus2012}.
As a consequence, the star spins up during the embedded phase and spins down afterward compared to a single star.
On the other hand, circumbinary disks have longer lifetimes compared to single stars \citep[e.g.,][]{Alexander2012}.
In multiple stellar systems, certain configurations can also result in long disk lifetimes \citep[e.g.,][]{Ronco2021}.

\subsection{Rotational distributions in young clusters}\label{sec:obs}

\begin{figure}
    \centering
         \resizebox{\hsize}{!}{\includegraphics{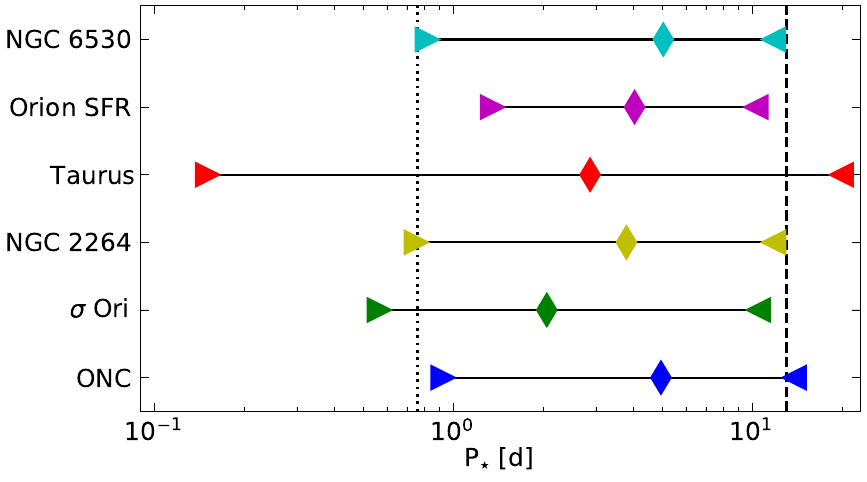}}
    \caption{
    Comparison between the rotational distribution in young clusters and rotation period range at 1~Myr of our fast and slow-rotating models (blue area).
    The diamonds, left-pointing triangles, and right-pointing triangles are the 50., 5., and 95.~percentile of the rotational distribution for the respective cluster. 
    The mean values of the 95. (5.)~percentiles are indicated by the vertical dashed (dotted) line.
    See text for references.
    }
    \label{fig:obsrot}
\end{figure}

Our results show the impact of different parameters on stellar spin evolution.
Now, we want to overview the observed rotational distributions of young clusters and estimate how fast (slow) a $\sim 1$~Myr old star probably spins.
In the next step, we infer the parameters necessary to reproduce the observed distribution and discuss the physical consequences of these choices (see \sref{sec:slowfast}).
We have selected the rotational period distribution of six young ($\lesssim 3$~Myr) clusters: NGC~6530 \citep[][]{Henderson12}, Orion~SFR \citep[][]{Serna2021}, Taurus \citep[][]{Rebull2020}, NGC~2264 \citep[][]{Venuti2017}, $\sigma$~Ori \citep[][]{Cody2010}, and ONC \citep[][]{Rebull2006}. 
In \fig{fig:obsrot}, we show the 5. (left-facing triangle), 50. (diamond), and 95. (right-facing triangle) percentile for each cluster.
For all six clusters, the 95.~percentiles indicating the slow-rotating stars are located at $\gtrsim 10$~days with a mean value of 13.0~days (vertical dashed line).
The fast-rotating boundary is at $\lesssim 1$~day with the exception of Taurus, which contains faster-rotating stars with the 5.~percentile at 0.15~days.
The mean value of the 5.~percentile is 0.76~days (vertical dotted line).
One explanation for the rapidly rotating stars in Taurus could be its age and the evolutionary stage of the star-disk systems.
With an age of $\sim 3$~Myr, the sample of Taurus consists of disk-bearing stars and stars without an accretion disk \citep[][]{Rebull2020}.
In addition, the ages of these young clusters are subjected to large scatters in the order of up to $\sim 1$~Myr \citep[e.g.,][]{Testi2022}. 
Some stars might be significantly older than the final age of our models (1~Myr).
During this additional time, the accretion disk could be dissolved due to accretion, winds, or photo-evaporative effects and the star starts to spin up due to contraction.
This theory is supported by a comparison of stars with and without an accretion disk in \cite{Rebull2020}.
Stars, which are still surrounded by an accretion disk, tend to rotate slower than stars without a disk.
The fastest, disk-bearing stars have rotation periods slightly below 1~day, which matches the 5.~percentiles of the other five clusters in our selection.

\subsection{Slow and fast rotating configurations}\label{sec:slowfast}

Based on the rotational distribution of young star-forming regions (see \sref{sec:obs}), 90\% of the stars in young clusters show rotation periods between 0.76~days and 13.0~days. 
While the rotation periods of our models match the fast-rotating stars in observed clusters (especially in the case of $\beta = 0.005$ and $W=0\%$, see \fig{fig:rot}), observed slow-rotating stars have rotation periods that are higher compared to our model results by a factor of $\gtrsim 4$.
We want to explore possible scenarios that can lead to these rotation periods.

A rising number of observed rotational distributions of young clusters allows for deriving empirical relations between the stellar rotation period and other parameters.
For accreting T~Tauri stars, a positive correlation between the rotation period and the stellar magnetic field strength has been reported \citep[e.g.,][]{Vidotto2014}.
For Class~II stars, a strong magnetic field result in a stellar spin down.
This relation agrees with our results (see Panel b in \fig{fig:rot}) and other studies \citep[e.g.,][]{Matt10, Ireland21, Gehrig2022}.
Since the magnetic field strength alone does not spin down the star sufficiently to match the observed slow-rotating stellar periods (see \fig{fig:rot}), the question arises if strong magnetic fields (in the star or the disk) might affect other parameters.
\cite{Bu2018} report that disk winds in a strongly magnetized environment are more powerful compared to a weakly magnetized outflow.
Similarly, \cite{Bai2016} explored the relation between angular momentum loss in protoplanetary disks due to the disk winds and the magnetic field strength.
The higher the magnetic field strength, the more effective is the wind in removing angular momentum from the disk and the higher the accretion rate induced by the wind.
A stronger accretion rate increases the amount of additional heat, $\Dot{E}_\mathrm{add}$, which is added to the star.
In addition, a strong magnetic field affects the accretion-powered stellar wind.
The Alfvén radius, $r_\mathrm{A}$, which acts as the lever arm of the APSW grows with an increasing stellar magnetic field strength \citep[e.g.,][]{Gallet19, Ireland21}.
The effect of the magnetic field strength on the mass-loss rate of the APSW, however, is poorly constrained by both, theory and observations.
To summarize, a strong magnetic field strength in the star and the disk can result in an effective disk wind, APSW, and can increase the accretion rate onto the star.
On the other hand, a weak magnetic field strength would be accompanied by weak winds (APSW and disk winds).

\begin{figure}
    \centering
         \resizebox{\hsize}{!}{\includegraphics{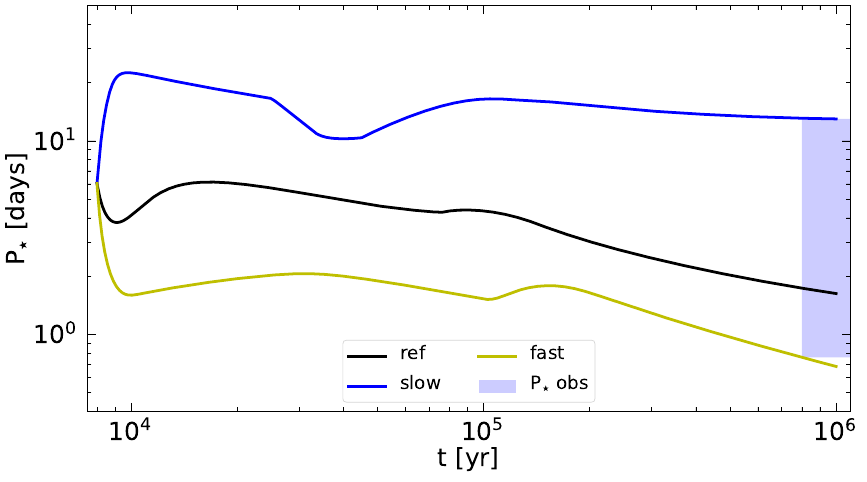}}
    \caption{
    Evolution of the stellar rotation period, $P_\star$ for the \texttt{slow} (blue line), reference (black line), and \texttt{fast} rotating models (yellow line).
    For comparison, the range of the mean values of the 5. and 95.~percentile is highlighted as a blue area, $P_\mathrm{\star~obs}$ (see \sref{sec:obs}).
    }
    \label{fig:slowfast}
\end{figure}

To test the effects of these configurations on the stellar spin evolution, two models \texttt{slow} and \texttt{fast} are constructed.
Model \texttt{slow} is characterized by a magnetic field strength of $B_\star = 2.0$~kG, the inclusion of disk winds removing angular momentum from the inner disk (see \sref{sec:disk_winds}), and $\Dot{M}_\mathrm{high}$ (see \sref{sec:change_eadd}).
All other values match the reference model.
Model \texttt{fast} combines a weak magnetic field strength, $B_\star=0.1$~kG, with an ineffective wind (no disk winds and $W=0\%$).
In \fig{fig:slowfast}, the evolution of the rotation periods for the model \texttt{slow} (blue line) and \texttt{fast} (yellow line) are shown in comparison to the reference model (black line).
The slow-rotating model spins down during the initial inflation to 22~days. 
After spinning up to $\sim 10$~days during the embedded phase, the strong magnetic field strength spins down the star, reaching 1~Myr with a period of 12.9~days.
The fast-rotating model spins up initially to a rotation period of 1.7~days due to the weak APSW wind.
This period is kept roughly constant during the embedded phase.
During the Class~II phase, the weak magnetic field strength results in a further spin-up to a rotation period of 0.68~days at an age of 1~Myr.
Thus, the effect of the magnetic fields in the young star-disk system on other parameters like the effectiveness of winds or the accretion rate can result in rotation periods that agree with the observed range in young clusters (blue shaded are in \fig{fig:slowfast}).
We note that the influence of magnetic fields on the amount of additional energy controlled by $\beta$ is unclear.
Additionally, the magnetic field strength itself is a free parameter in our model.
The generation of the magnetic field and its evolution is still unclear.
One possible evolution scenario for the magnetic field strength is given by \cite{Browning16}.
The magnetic energy is scaled with the convective energy in the stellar interior and a positive correlation between the stellar mass and the magnetic field strength is proposed.
Further investigation is needed to constrain the evolution of the free parameters used in our model.

Another empirical relation connects the stellar rotation period and the stellar mass.
In \cite{Herbst2007}, a higher stellar mass is associated with longer rotation periods in $\sim 1$~Myr old clusters.
We note that in the context of this work, a high-mass star corresponds to a solar-mass star.
On the other hand, a weak and even negative correlation between the stellar rotation period and the stellar mass is reported in \cite{Henderson12} for cluster ages of $\sim 1$~Myr.
Our results presented in \fig{fig:rot} agree with the negative correlation between stellar period and mass.
However, the disks surrounding high-mass stars, which are more massive compared to low-mass stars \citep[e.g.,][]{Vorobyov2011, Andrews2013}, could launch stronger disk winds as these winds scale with the density in the disk \citep[e.g.,][]{Bai2016}.
In addition, the stellar mass scales with the accretion rate \citep[e.g.,][]{Testi2022, Somigliana2022} and a high-mass star could be injected with a larger amount of additional heat, $\Dot{E}_\mathrm{add}$.
Thus, a combination of high stellar masses, strong accretion rates, and an effective disk wind can also result in a slow-rotating star.
To test the relation between the stellar mass and the rotation period in more detail and constrain the mentioned parameters, a combined stellar and disk evolution model, including disk winds, and further observations are required.

\section{Conclusion}
\label{sec:conclusion}

In this work, we present the evolution of stellar rotation periods for young ($\leq 1$~Myr), low-mass ($\leq 1~\mathrm{M_\odot}$) stars.
We start our simulation from a stellar seed ($0.05~\mathrm{M_\odot}$) and take protostellar accretion, stellar winds, the magnetic star-disk interaction as well as the internal stellar evolution into account. 
Our model allows us to test the influence of a variety of parameters on the stellar rotation period.
Based on our results, we can draw the following conclusions:

\begin{itemize}

    \item 
    The stellar rotation period is significantly influenced by the amount of additional heat, $\Dot{E}_\mathrm{add}$, (controlled by $\beta_\mathrm{up}$) that is injected into the stellar interior, the accretion history, and the presence of strong winds (APSW and disk winds). 
    The magnetic field strength, which is dominant during later evolutionary stages, plays a noticeable but smaller role during the first million years.

    \item
    Differences in the initial rotational periods of the stellar seed, $\Omega_\mathrm{init}$, are quickly forgotten ($\lesssim 10^4$~years) and do not impact the rotational distribution at later stellar ages. 
    Similarly, the initial stellar radius, $R_\mathrm{init}$, (assuming $\beta_\mathrm{up}\sim 0.2$) and the extent of the region, in which the additional heat is distributed, $M_\mathrm{add}$, do not affect the rotation period noticeably.

    \item
    We explore the possibility of fast and slow-rotating models.
    The magnetic field strength in the star-disk system can affect the amount of additional heat, $\Dot{E}_\mathrm{add}$, and the effectiveness of both, stellar and disk winds.
    A strong magnetic field can lead to an enhanced accretion rate and effective winds (APSW and disk winds), resulting in slow-rotating stars.
    Weak magnetic fields or a small amount of additional heat ($\beta = 0.005$) lead to the fastest rotating stars.
    The stellar rotation periods in our models cover a range from $0.6-12.9$~days.

    \item 
    Our models match the rotational distribution of six young ($\lesssim 3$~Myr) clusters. 
    The rotation periods of up to 90\% of all observed stars are located within our results.

    \item
    Motivated by these intriguing results, our model, extended over the whole disk lifetime, can be used as an alternative to the widely used disk-locking model during the presence of an accretion disk.
    In combination with a hydrodynamic disk model, including effects such as external photo-evaporation and episodic accretion outbursts, valuable insights into the origin of the rotational distribution of young clusters can be obtained.

\end{itemize}

\vspace{5mm}


\begin{acknowledgements}
We thank the anonymous referee for the constructive feedback that helped to improve the manuscript.
The authors gratefully acknowledge the work of Bill Paxton and his collaborators on the stellar evolution code MESA.
Furthermore, the authors thank Thomas Steindl and Konstanze Zwintz for providing valuable insights into "handling" MESA and its parameter files.
E.I.V. acknowledges support by the Ministry of Science and Higher Education of the Russian Federation (State assignment in the field of scientific activity 2023).
The simulations were performed on the VSC Vienna Scientific Cluster.
 
\end{acknowledgements}

\bibliographystyle{resources/bibtex/aa}
\bibliography{literature/Tapire, literature/Rodrigo}
%


\begin{appendix}
\section{MESA microphysics}
\label{app:mesaphysics}
The MESA EOS is a blend of the OPAL \citet{Rogers2002}, SCVH
\citet{Saumon1995}, PTEH \citet{Pols1995}, HELM
\citet{Timmes2000}, and PC \citet{Potekhin2010} EOSes.

Radiative opacities are primarily from OPAL \citep{Iglesias1993,
Iglesias1996}, with low-temperature data from \citet{Ferguson2005}
and the high-temperature, Compton-scattering-dominated regime by
\citet{Buchler1976}.  Electron conduction opacities are from
\citet{Cassisi2007}.

Nuclear reaction rates are a combination of rates from
NACRE \citep{Angulo1999}, JINA REACLIB \citep{Cyburt2010}, plus
additional tabulated weak reaction rates \citet{Fuller1985, Oda1994,
Langanke2000}. Screening
is included via the prescription of \citet{Chugunov2007}.  Thermal
neutrino loss rates are from \citet{Itoh1996}.

\FloatBarrier
\section{Treatment of rotation in MESA}
\label{app:mesarot}

In MESA, rotation is treated as a modification of the stellar equations due to centrifugal acceleration and non-spherical symmetry, when the star is rotating \citep[see Sec. 6 in][]{Paxton2013}.
Angular momentum transport due to rotation (rotational mixing) is implemented as a diffusion term, including five rotational-induced mixing processes: dynamical shear instability, Solberg–Høiland instability, secular shear instability, Eddington–Sweet circulation, and the Goldreich–Schubert–Fricke instability \citep[][]{Paxton2013}.

\FloatBarrier

\section{Detailed evolution of individual torque components}\label{sec:tor_comp}

We show each component affecting the stellar spin evolution: the external torques due to the accretion process $\Gamma_\mathrm{acc}$, APSW $\Gamma_\mathrm{W}$ and the magnetic star-disk interaction $\Gamma_\mathrm{ME}$, the internal contribution $\Gamma_\mathrm{int}$ and the resulting total torque $\Gamma_\mathrm{tot}$.

\begin{figure}
    \centering
         \resizebox{\hsize}{!}{\includegraphics{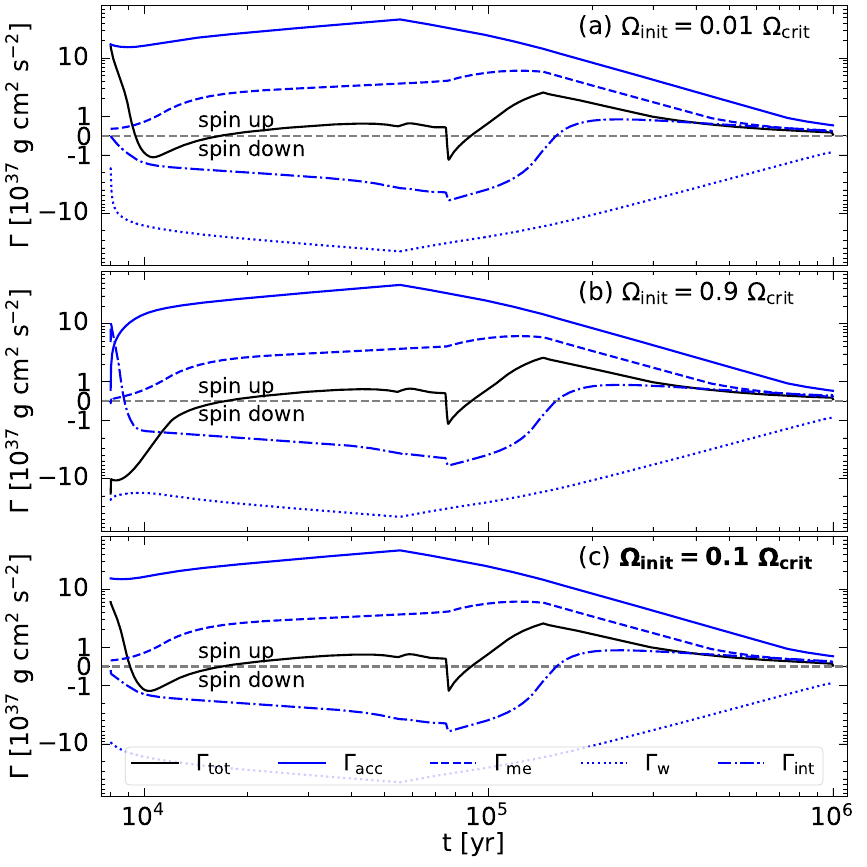}}
    \caption{
    Evolution of the different external torque components ($\Gamma_\mathrm{acc}$, $\Gamma_\mathrm{me}$ and $\Gamma_\mathrm{w}$; solid blue, dashed blue, and dotted blue lines, respectively), the internal contribution to the stellar spin evolution $\Gamma_\mathrm{int}$ (blue dash-dotted line) and the combination of these mechanisms $\Gamma_\mathrm{tot}$ (black solid line) for different initial rotation rates, $\Omega_\mathrm{init}$.
    The value of the reference model is marked in boldface.
    The horizontal dashed black line divides the regions, in which the star spins up and down. 
    }
    \label{fig:tor_omega}
\end{figure}

\begin{figure}
    \centering
         \resizebox{\hsize}{!}{\includegraphics{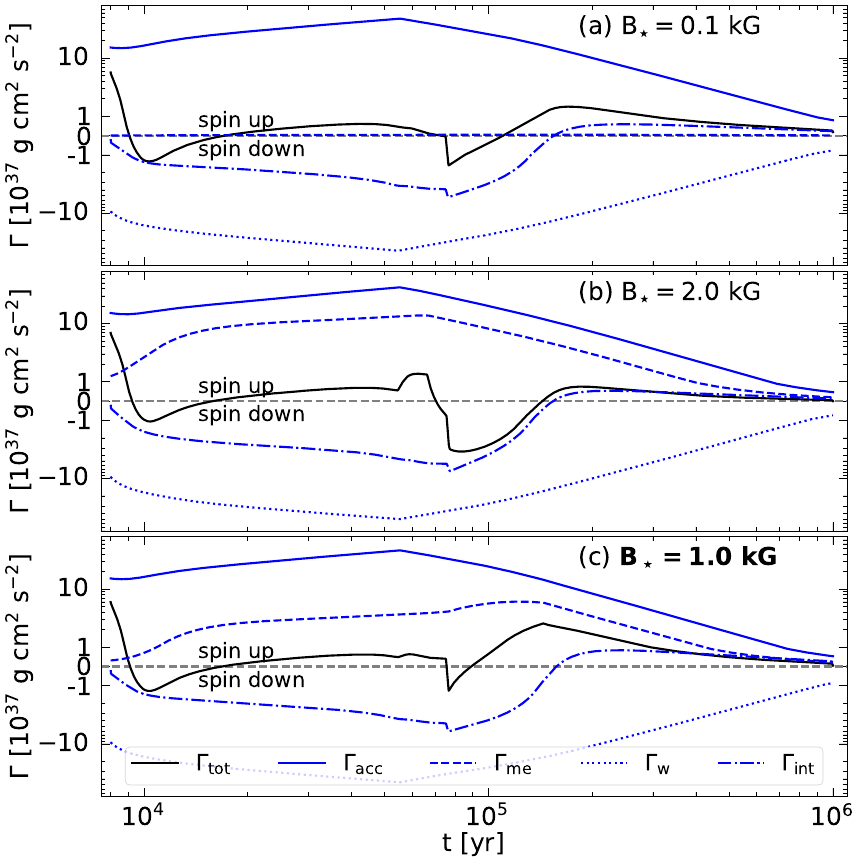}}
    \caption{
    Same as \fig{fig:tor_omega} for different magnetic field strengths, $B_\star$.
    }
    \label{fig:tor_bstar}
\end{figure}

\begin{figure}
    \centering
         \resizebox{\hsize}{!}{\includegraphics{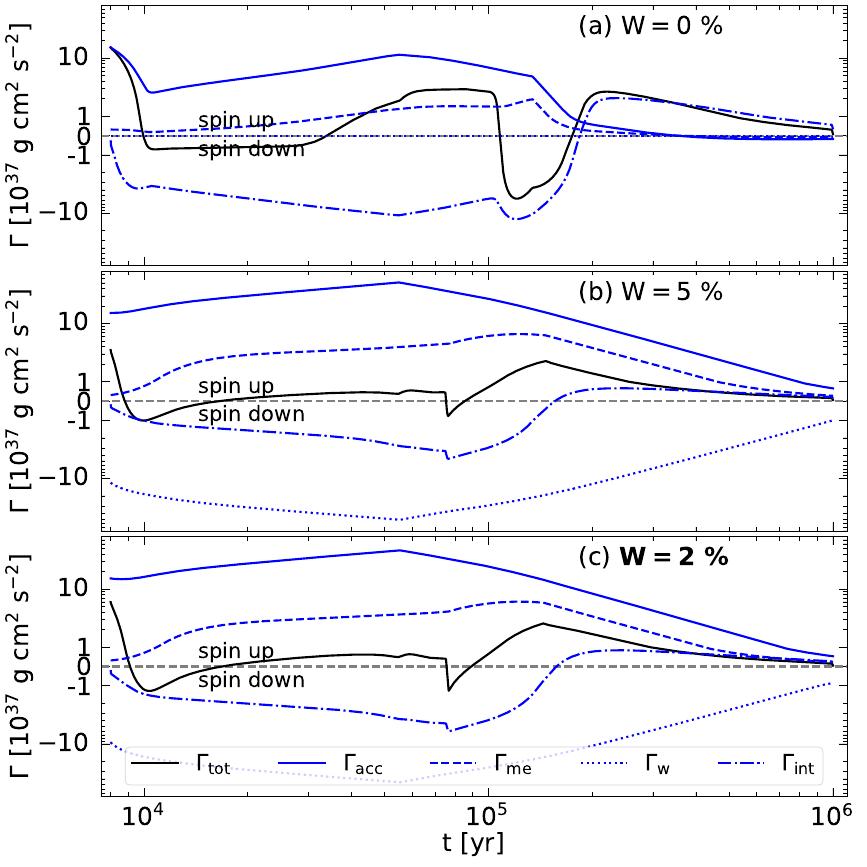}}
    \caption{
    Same as \fig{fig:tor_omega} for different APSW efficiencies,$W$.
    }
    \label{fig:tor_apsw}
\end{figure}

\begin{figure}
    \centering
         \resizebox{\hsize}{!}{\includegraphics{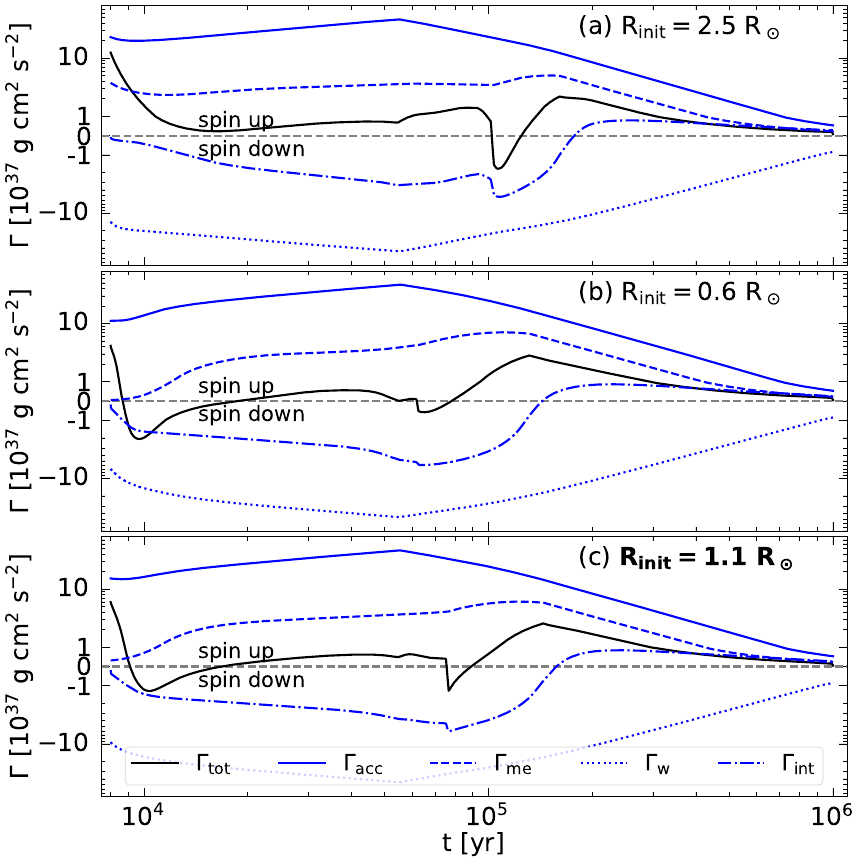}}
    \caption{
    Same as \fig{fig:tor_omega} for different initial stellar radii, $R_\mathrm{init}$.
    }
    \label{fig:tor_rinit}
\end{figure}

\begin{figure}
    \centering
         \resizebox{\hsize}{!}{\includegraphics{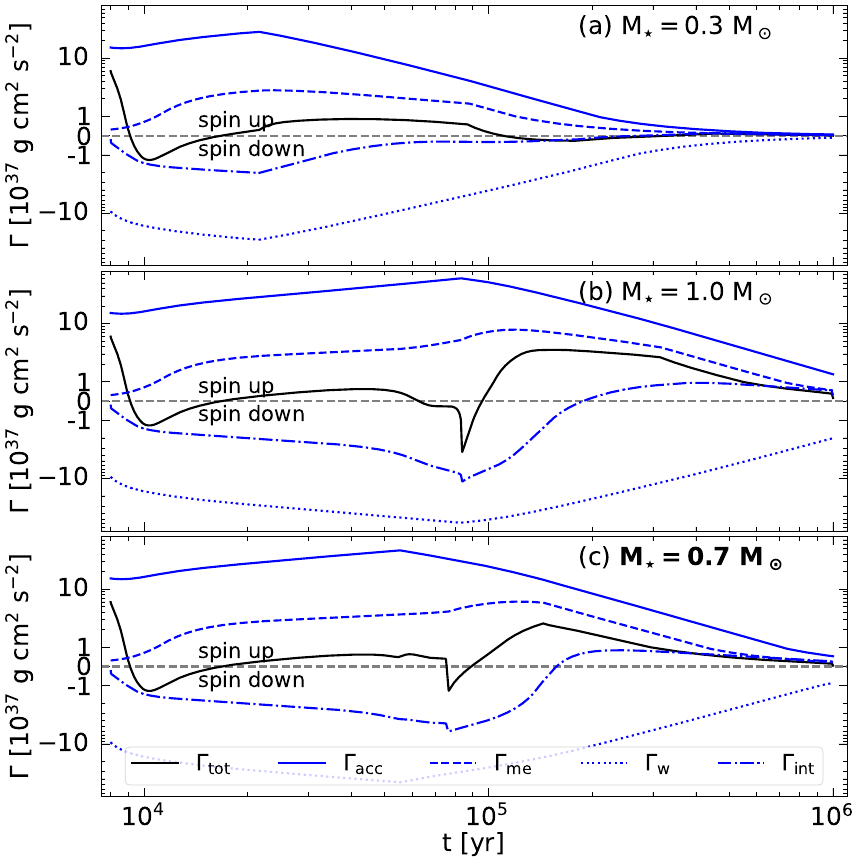}}
    \caption{
    Same as \fig{fig:tor_omega} for different stellar masses, $M_\star$, at an age of 1~Myr.
    }
    \label{fig:tor_mstar}
\end{figure}

\begin{figure}
    \centering
         \resizebox{\hsize}{!}{\includegraphics{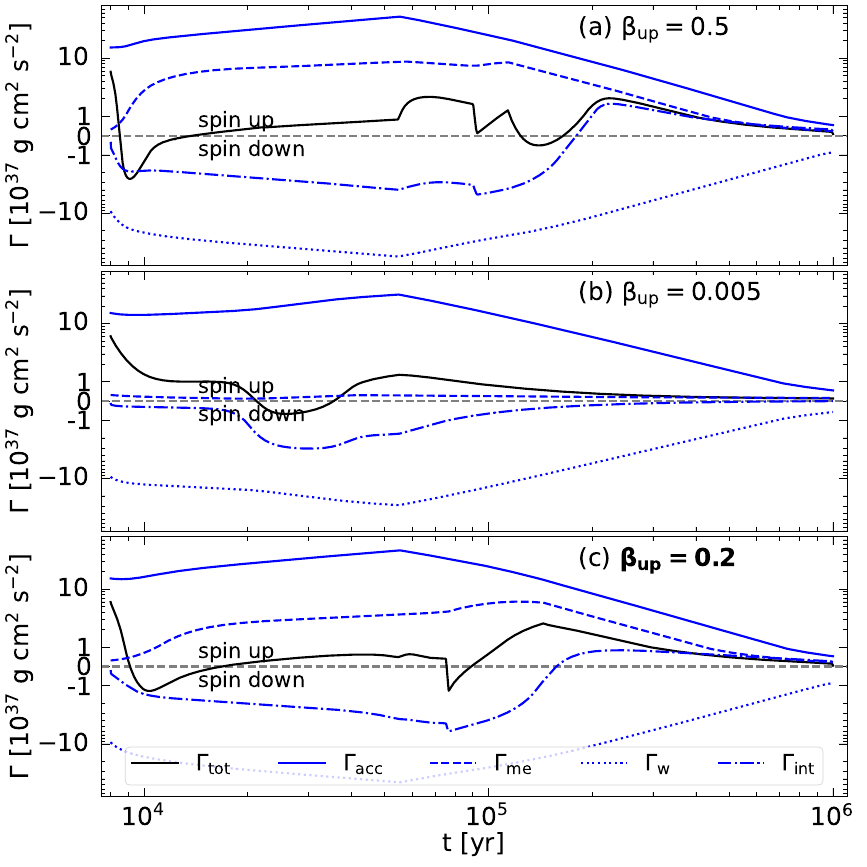}}
    \caption{
    Same as \fig{fig:tor_omega} for different maximum energy fractions that can be added to the stellar interior, $\beta_\mathrm{up}$.
    }
    \label{fig:tor_beta}
\end{figure}

\FloatBarrier

\section{Influence of the initial stellar radius for $\beta_\mathrm{up}=0.005$}\label{sec:rin_beta}

Evolution of the stellar rotation period and radius for $\beta_\mathrm{up}=0.005$ and different initial stellar radii.

\begin{figure}[ht!]
    \centering
         \resizebox{\hsize}{!}{\includegraphics{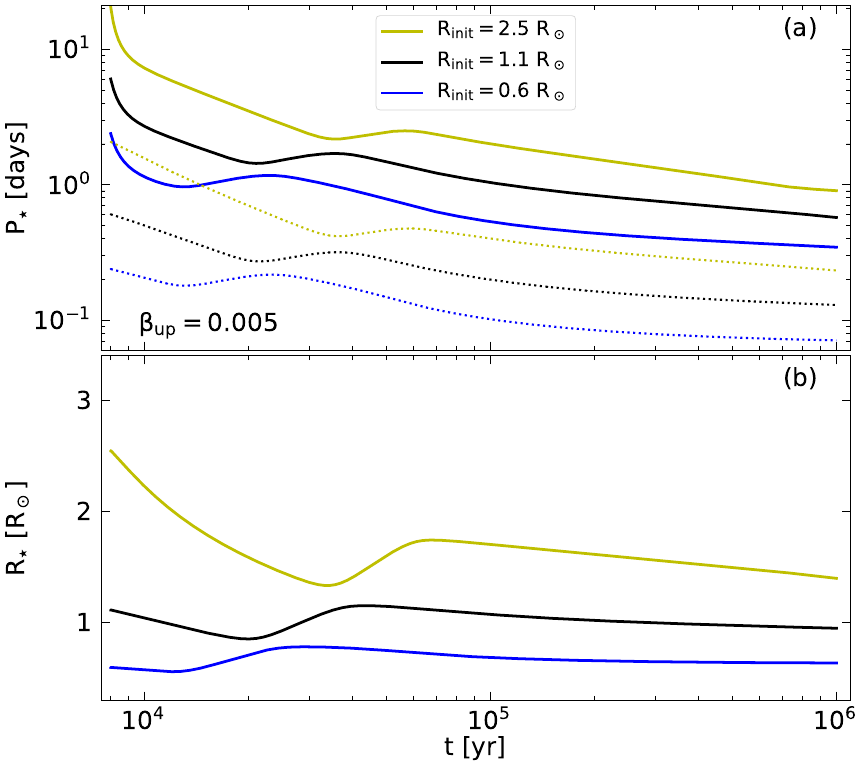}}
    \caption{
    Evolution of the stellar rotation periods, $P_\star$ (Panel a), and the stellar radius (Panel b) for different initial stellar radii, $R_\mathrm{init}$.
    For these simulations we choose $\beta_\mathrm{up}=0.005$.
    }
    \label{fig:rin_beta}
\end{figure}

\FloatBarrier

\section{HRD for individual simulations}\label{sec:hrd}

Here, we show the influence of each parameter on the HRD.

\begin{figure*}[ht!]
    \centering
         \resizebox{\hsize}{!}{\includegraphics{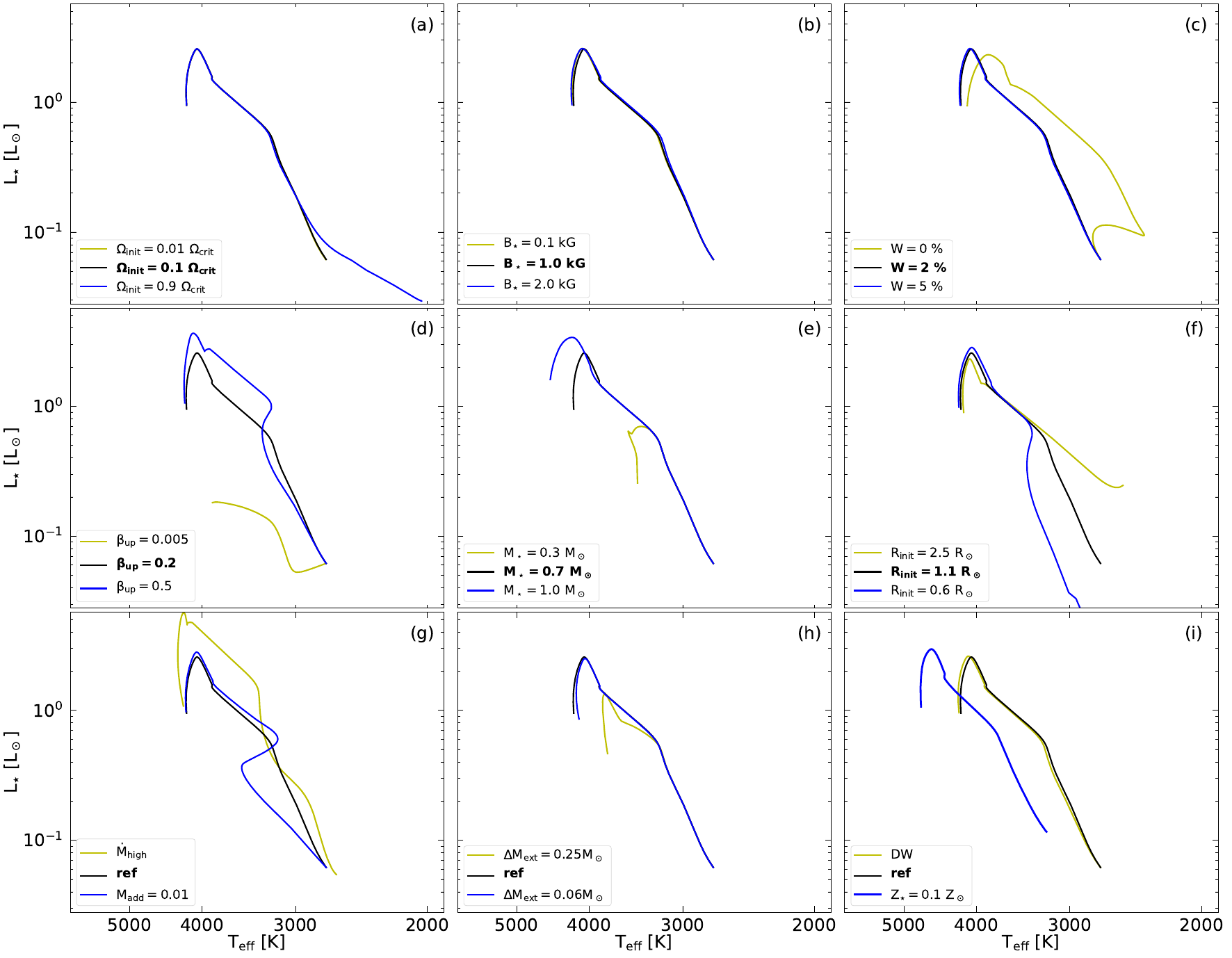}}
    \caption{
    Intrinsic stellar luminosity, $L_\star$, over effective temperature, $T_\mathrm{eff}$, for each model presented in \sref{sec:results} and \sref{sec:discussion}.
    The reference model is marked in boldface.
    }
    \label{fig:hrd}
\end{figure*}

\FloatBarrier

\end{appendix}

\end{document}